\documentclass[pdflatex,prd,twocolumn,showpacs,superscriptaddress,nofootinbib]{revtex4-1}
\usepackage{bm,amsmath,amssymb}
\usepackage{graphicx}
\usepackage{subfigure}
\usepackage{color}
\usepackage{soul}
\usepackage{hyperref}
\usepackage{multirow}
\usepackage{lineno}
\usepackage{hyperref}

\hypersetup{
    colorlinks = true,
    linkcolor = blue,
    anchorcolor = blue,
    citecolor = blue,
    filecolor = blue,
    urlcolor = blue
    }
\setulcolor{blue}
\setstcolor{red}
\sethlcolor{yellow}
\def\lsim{\raise0.3ex\hbox{$<$\kern-0.75em\raise-1.1ex\hbox{$\sim$}}}
\def\gsim{\raise0.3ex\hbox{$>$\kern-0.75em\raise-1.1ex\hbox{$\sim$}}}

\newcommand \beq {\begin{equation}}
\newcommand \eeq {\end{equation}}
\newcommand \beqa {\begin{eqnarray}}
\newcommand \eeqa {\end{eqnarray}}
\newcommand \hmu {\hat{\mu}}
\newcommand \xb {\bar{x}}
\newcommand \cb {\bar{\chi}}

\newcommand{\muQ}{\mu_Q}

\newcommand{\pade}{Pad\'e}
\begin{document}
%\linenumbers

\title{Taylor expansions and Pad\'e approximants
for cumulants of conserved charge fluctuations 
at non-vanishing chemical potentials}

\author{D. Bollweg}
\affiliation{Physics Department, Columbia University, New York, NY 10027, USA}
\author{J. Goswami}
\affiliation{Fakult\"at f\"ur Physik, Universit\"at Bielefeld, D-33615 Bielefeld, Germany}
\author{O. Kaczmarek}
\affiliation{Fakult\"at f\"ur Physik, Universit\"at Bielefeld, D-33615 Bielefeld, Germany}
\author{F. Karsch}
\affiliation{Fakult\"at f\"ur Physik, Universit\"at Bielefeld, D-33615 Bielefeld, Germany}
\author{Swagato Mukherjee}
\affiliation{Physics Department, Brookhaven National Laboratory, Upton, New York 11973, USA}
\author{P. Petreczky}
\affiliation{Physics Department, Brookhaven National Laboratory, Upton, New York 11973, USA}
\author{C. Schmidt}
\affiliation{Fakult\"at f\"ur Physik, Universit\"at Bielefeld, D-33615 Bielefeld, Germany}
\author{P. Scior}
\affiliation{Physics Department, Brookhaven National Laboratory, Upton, New York 11973, USA}

\collaboration{HotQCD collaboration}
\date{\today}

\begin{abstract}
Using high statistics datasets generated in (2+1)-flavor QCD calculations at finite temperature we present results for low order cumulants of net
baryon-number fluctuations at 
non-zero values of the baryon chemical potential.
We calculate Taylor expansions for the pressure
(zeroth order cumulant), net baryon-number density (first order cumulant) and the variance of the distribution on net-baryon number
fluctuations (second order cumulant). We obtain series expansions from an eighth order expansion of the pressure
and compare these to diagonal Pad\'e approximants.
This allows us to estimate the range of values for the baryon chemical potential in which these 
expansions are reliable. We find $\mu_B/T\le 2.5$, $2.0$ and $1.5$ for the zeroth, first
and second order cumulants, respectively. We furthermore, construct estimators for
the radius of convergence of the Taylor series of the pressure. In the vicinity of the pseudo-critical temperature, $T_{pc}\simeq 156.5$~MeV, we find $\mu_B/T\ \gsim\ 2.9$ at vanishing strangeness chemical potential and somewhat larger values for strangeness neutral matter. These estimates are temperature dependent and range from $\mu_B/T\ \gsim\ 2.2$ at $T=135$~MeV to 
$\mu_B/T\ \gsim\ 3.2$ at $T=165$~MeV. The estimated radius of convergences is the same for any higher order cumulant.

\end{abstract}


\maketitle

\section{Introduction}

While we gained a lot of information on the thermodynamics of strong interaction matter through
numerical calculations in the framework of lattice
regularized Quantum Chromodynamics (QCD) at finite 
temperature, the extension to non-vanishing values of
conserved charge densities, {\it i.e.} net baryon-number ($B$), electric charge ($Q$) and strangeness ($S$), 
is difficult due to the lack of appropriate numerical techniques. The currently most actively pursued approaches to QCD at non-zero temperature and non-zero
conserved charge densities are based on the application of 
Taylor series expansions in terms of conserved charge
chemical potentials, $\vec{\mu}=(\mu_B, \mu_Q, \mu_S)$ \cite{Gavai:2004sd,Allton:2005gk},
or direct simulations at non-zero imaginary chemical potentials \cite{DElia:2002tig,Bellwied:2015rza} (for recent reviews see e.g. \cite{Ding:2015ona,DElia:2018fjp,Guenther:2022wcr}). While the former 
approach has to deal with the range of validity of series expansions arising
from a finite radius of convergence of such expansions and truncation errors arising from the limited knowledge
on the number of expansion parameters
\cite{Karsch:2010hm}, the latter requires analytic continuation to physical, real
values of the chemical potential and thus is limited 
by the ansatz used for analytic continuation of thermodynamic observables \cite{Guenther:2017hnx,Bonati:2018nut}, which in practice also is 
limited by the statistical accuracy with which parameters of such an analytic continuation can be constrained.

Recently much effort has been put into a better understanding of the analytic structure of the 
QCD partition function as function of complex valued
chemical potentials. This is important for our 
ability to generate suitable ans\"atze for the 
analytic continuation of calculations performed
with imaginary values of the chemical 
potentials as well as for choosing appropriate
resummation schemes that allow to extend results obtained in Taylor series beyond the radius of convergence of such expansions.
Poles of the logarithm of the QCD partition function in the complex chemical potential plane might be of 
simple thermal origin, arising \textit{e.g.} from the 
analytic structure of Fermi or Bose distribution functions \cite{Skokov:2010uc}, or stem from universal critical behavior, known as Lee-Yang and Lee-Yang edge singularities \cite{Stephanov:2006dn,Pradeep:2019ccv, Basar:2021hdf,Mukherjee:2021tyg}. Studies of Lee-Yang zeros/singularities have a long history in QCD, recent studies include \textit{e.g.}, \cite{Wakayama:2018wkc,Giordano:2019gev,Mondal:2021jxk}. The scaling of the Lee-Yang edge singularities and its influence on the QCD phase transition was considered only recently \cite{Mukherjee:2019eou,Basar:2021hdf,Nicotra:2021ijp,Dimopoulos:2021vrk,Schmidt:2021pey}. 

We will focus here on the analysis of the Taylor series
expansion of the partition function of (2+1)-flavor 
QCD and discuss the resummation of such series using
Pad\'e approximants.
The range of validity of Taylor expansions using 
cumulants calculated at physical values of the quark masses
is limited by singularities of the logarithm of the QCD partition function, {\it i.e.} the pressure, that occur for complex valued $\vec{\mu}$. These singularities may either
occur for real values of  $\vec{\mu}$ or in the
complex plane, e.g. where ${\rm Im}( \mu_B) \ne 0$. Only poles on the real $\vec{\mu}$-axis 
correspond to phase transitions in QCD. As recent studies of (2+1)-flavor QCD with lighter than physical quark masses have shown that the chiral phase transition temperature is 
at $T_c=132^{+3}_{-6}$~MeV \cite{HotQCD:2019xnw} and as this 
is expected to set an upper bound on the location of a possible critical point at non-zero values of the baryon chemical potential \cite{Halasz:1998qr}, we expect to
find only complex poles for the analytically continued pressure in (2+1)-flavor 
QCD at temperatures above $T\simeq 135$~MeV. Such singularities will limit the radius of convergence for the 
Taylor series, which has been estimated ever since the first 
applications of the Taylor expansion approach in lattice QCD calculations \cite{Gavai:2004sd,Allton:2005gk,Giordano:2019gev}. 

The singularities in QCD
partition functions in the complex $\mu_B$-plane also
have impact on the range of applicability of series
expansions performed at real values of the chemical 
potentials.
Limitations for the determination of the searched for critical point in QCD, arising from a finite 
radius of convergence of Taylor expansions, can however be circumvented by using appropriate resummation schemes for the Taylor series \cite{Gavai:2008zr,Skokov:2010uc,Datta:2016ukp,Vovchenko:2017gkg,Mondal:2021jxk,Borsanyi:2021sxv,Borsanyi:2022qlh}. Using Pad\'e
approximants is one way to gain information on the analytic structure of the QCD partition function. They allow to explore e.g. the 
pressure of QCD beyond the limit set by a finite radius of convergence of Taylor series \cite{Karsch:2010hm,Dimopoulos:2021vrk,Schmidt:2021pey,Datta:2016ukp, Basar:2021gyi}.

Results for Taylor expansion coefficients, {\it i.e.} the cumulants $\chi_{ijk}^{BQS}$, in (2+1)-flavor QCD 
up to $8^{\rm th}$ order, {\it i.e.} for all $0 <(i+j+k) \le 8$, get 
improved steadily by the HotQCD Collaboration \cite{HotQCD:2018pds,Bazavov:2020bjn,Bollweg:2021vqf} in calculations with the 
Highly Improved Staggered Quark (HISQ) action \cite{Follana:2006rc}.
These expansion coefficients have been used
for a determination of the line of pseudo-critical
temperatures $T_{pc}(\mu_B)$ \cite{HotQCD:2018pds}
and in an analysis of high order cumulants at 
non-vanishing values of the chemical potentials 
\cite{Bazavov:2020bjn}. The datasets used in these
calculations have been extended 
by adding calculations at a lower temperature, 
$T\simeq 125$~MeV,  for lattices with temporal extent $N_\tau=8$, and more statistics has been added 
on lattices with temporal extent $N_\tau=12$ and $16$.
Based on these updated
datasets we presented in Ref.~\cite{Bollweg:2021vqf} an analysis of
first and second order cumulants at vanishing values of the chemical
potentials. Using in addition the results for higher order cumulants we present
here an analysis of the low order cumulants at non-vanishing
values of the chemical potentials. 
The datasets that are now available for Taylor coefficients
calculated with the HISQ action 
contain more than a factor 10 higher statistics on lattices with 
temporal extent $N_\tau=8$ and a factor 20 higher statistics for 
$N_\tau=12$  than used previously in studies at non-zero $\vec{\mu}$.  This allows a careful analysis of
the reliability range of such expansions on the one hand and an estimate of the radius of convergence of the
Taylor series on the other hand.

Aside from a systematic discussion of convergence properties of the Taylor series and their improvement through resummation of the 
series, applying techniques commonly used for other statistical
systems, e.g. Pad\'e approximations,
the analysis of low order 
cumulants also provides the basis for a new, highly improved, 
analysis of the QCD equation of state of $(2+1)$-flavor QCD at non-vanishing chemical potentials. 
We provide here results on the pressure and 
net baryon-number density and leave the analysis
of other bulk thermodynamic observables to a forthcoming publication.

This paper is organized as follows. In section II 
we present our results on Taylor expansions for 
the pressure of (2+1)-flavor QCD, the net baryon-number density, and
the second order cumulant of net baryon-number fluctuations, involving cumulants of up to eighth order. In Section III we construct
Pad\'e approximants for these Taylor series
and discuss information on the location of poles closest to the origin that give estimators for the radius of convergence of the Taylor series. Section IV presents a comparison of Taylor series and Pad\'e approximants that allows to estimate the range of chemical potentials in which current series expansions, that can be 
constructed by using up to $8^{th}$ order cumulants only, provide reliable results. Finally we give our conclusions in section V. In three Appendices we
present (A) an explicit expression for the eighth order Taylor expansion coefficient of the pressure, (B) additional expansion coefficients needed for the Taylor series of the second order cumulant of net baryon-number fluctuations in strangeness neutral, isospin symmetric systems and (C)
some details on poles of the diagonal [4,4] Pad\'e for the pressure in (2+1)-flavor QCD.

\section{Taylor expansions of low order cumulants and the equation of state}

\subsection{Computational set-up for Taylor expansion in (2+1)-flavor QCD}
The framework for our calculations with the HISQ \cite{Follana:2006rc} discretization scheme for $(2+1)$-flavor 
QCD with a physical strange quark mass and two degenerate, physical light quark masses
is well established and has been used by us in several studies of higher order
cumulants of conserved charge fluctuations and correlations. The specific set-up
used in our current study has been described in \cite{Bazavov:2020bjn}. The framework for Taylor series expansions for strangeness neutral systems with fixed ratio of net electric charge
to net baryon-number has been given
up to $6^{\rm th}$ order in Ref.~\cite{Bazavov:2017dus}. It has been extended
in Ref.~\cite{Bazavov:2020bjn} by providing the 
necessary expansion coefficients for calculations involving up to $8^{\rm th}$ order cumulants. In that publication 
the $8^{\rm th}$ order expansion coefficient  of the pressure in strangeness neutral systems was not included. In this work, we present an explicit expression for it in Appendix~\ref{app:Pn}.

\subsection{Taylor expansion coefficients up to and including
\boldmath ${\cal O}(\mu_B^8)$}

We consider thermodynamic quantities, in particular low order cumulants of conserved
charge fluctuations, derived from Taylor expansions for the 
pressure of $(2+1)$-flavor QCD,
\begin{equation}
	\frac{P}{T^4} = \frac{1}{VT^3}\ln\mathcal{Z}(T,V,\vec{\mu}) = \sum_{i,j,k=0}^\infty%
\frac{\chi_{ijk}^{BQS}}{i!j!\,k!} \hmu_B^i \hmu_Q^j \hmu_S^k \; ,
\label{Pdefinition}
\end{equation}
with $\hat{\mu}_X\equiv \mu_X/T$ and arbitrary natural numbers $i,j,k$.
The expansion coefficients, $\dfrac{\chi_{ijk}^{BQS}}{i!j!k!}$ are 
derivatives of $P/T^4$ with respect to the associated chemical potentials,
$\vec{\mu}=(\mu_B, \mu_Q, \mu_S)$, evaluated at
$\vec{\mu}=\vec{0}$,
\begin{equation}
\chi_{ijk}^{BQS} =\left. 
\frac{1}{VT^3}\frac{\partial \ln\mathcal{Z}(T,V,\vec{\mu}) }{\partial\hmu_B^i \partial\hmu_Q^j \partial\hmu_S^k}\right|_{\vec{\mu}=0} \; ,
\; i+j+k\; {\rm even} \; .
\label{suscept}
\end{equation}

\begin{figure*}[ht]
\includegraphics[scale=0.54]{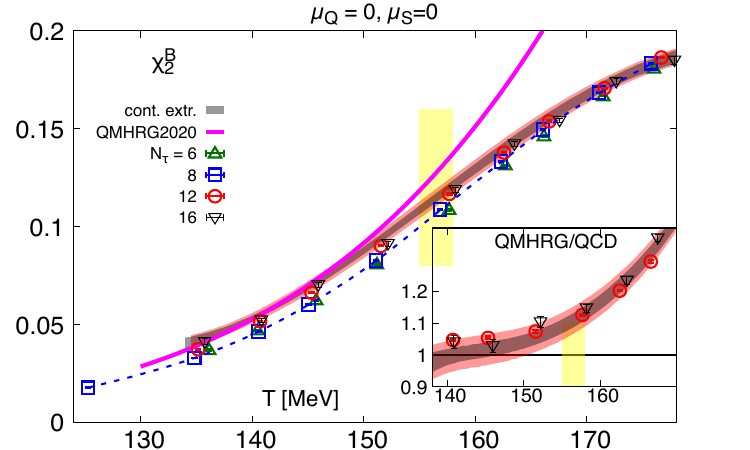}
\includegraphics[scale=0.54]{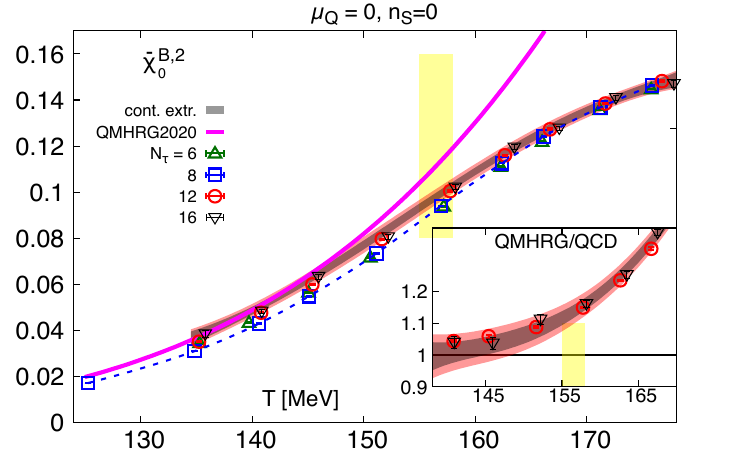}
\includegraphics[scale=0.54]{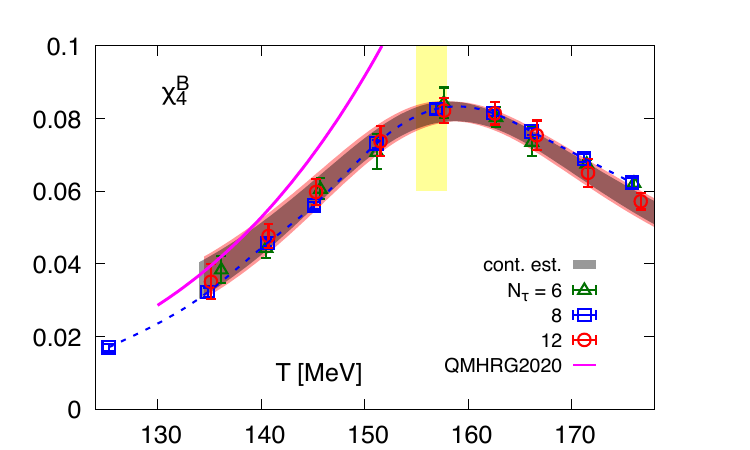}
\includegraphics[scale=0.54]{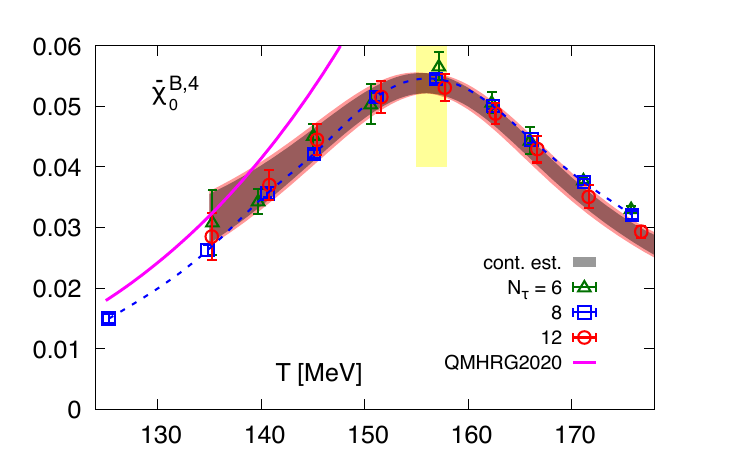}
\includegraphics[scale=0.54]{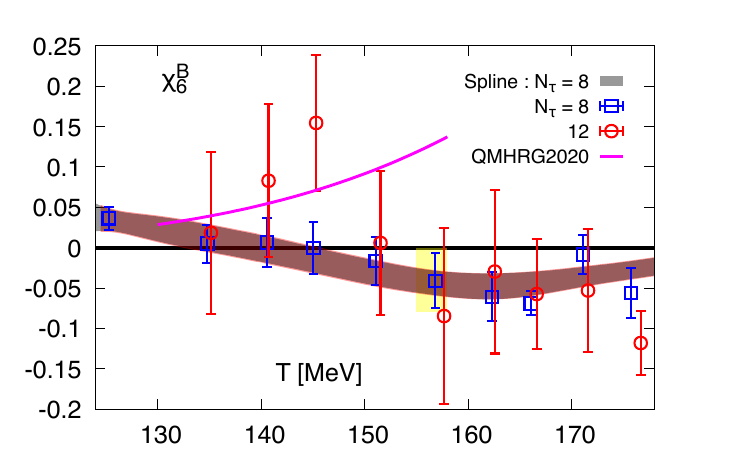}
\includegraphics[scale=0.54]{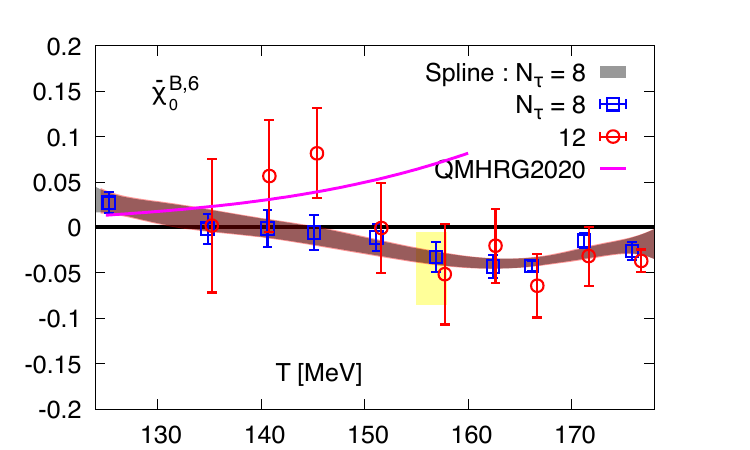}
\includegraphics[scale=0.54]{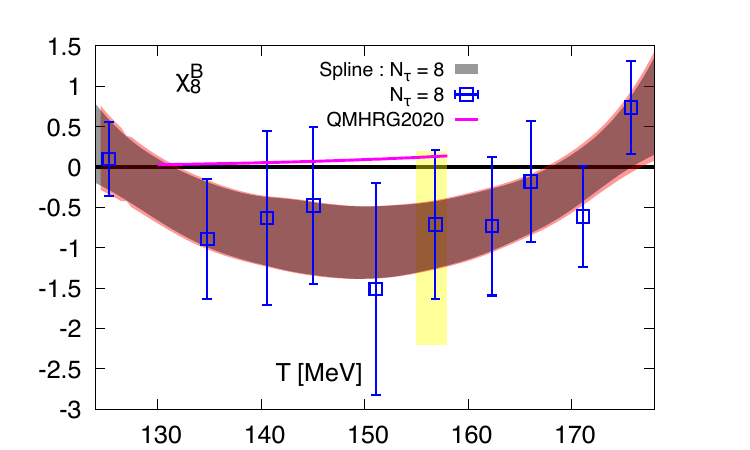}
\includegraphics[scale=0.54]{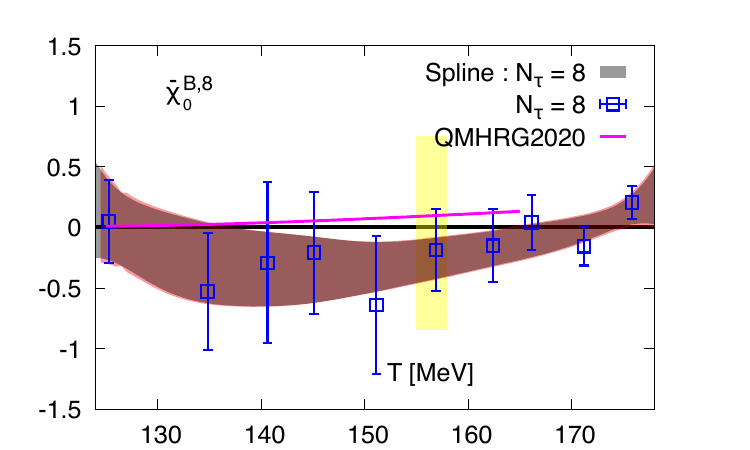}
\caption{The $n^{\rm th}$ order cumulants, $\cb_0^{B,n}$, contributing to the Taylor series of the pressure of
(2+1)-flavor QCD  as function of $\hmu_B=\mu_B/T$ versus temperature.
Shown are the expansion coefficients for the cases of (i) $\mu_Q=\mu_S=0$ (left column)
and (ii) $\mu_Q=0$, $n_S=0$ (right
column), respectively. In both cases the actual 
$n^{\rm th}$ order expansion coefficients in the Taylor series are obtained with these cumulants as $\cb_0^{B,n}/n!$. Yellow bands show the location of 
the pseudo-critical temperature $T_{pc}(0)=156.5(1.5)$~MeV \cite{HotQCD:2018pds}.
}
\label{fig:taylor8}
\end{figure*}

Aside from the Taylor expansion of the pressure
we will focus here on the analysis of Taylor
series for the first and second order cumulants
of net baryon-number fluctuations,
\begin{eqnarray}
n_B &=& \frac{\partial P/T^4}{\partial \hmu_B }
\;\;  , \label{nX} \\
\chi_2^B &=& \frac{\partial^2 P/T^4}{\partial \hmu_B^2 }  \;\; .
\label{chiX}
\end{eqnarray}
For these observables we will introduce
constraints on the electric charge and strangeness chemical potentials, \cite{Bazavov:2017dus,Bazavov:2020bjn} 
\begin{eqnarray}
        \hat{\mu}_Q(T,\mu_B) &=& q_1(T)\hat{\mu}_B + q_3(T) \hat{\mu}_B^3+q_5(T) \hat{\mu}_B^5+ \dots \; , \nonumber \\
        \hat{\mu}_S(T,\mu_B) &=& s_1(T)\hat{\mu}_B + s_3(T) \hat{\mu}_B^3+s_5(T) \hat{\mu}_B^5 +..  
\label{qs}
\end{eqnarray}
that enforce
strangeness neutrality ($n_S=0$) and a fixed
ratio $n_Q/n_B=r$. Here the case $r=0.5$ refers to an isospin symmetric medium, which is realized for $\mu_Q=0$. The case $r=0.4$ corresponds to the 
situation met in heavy ion collision experiments.
Explicit expressions for the expansion coefficients $q_i$ and $s_i$ with $i=1,\ 3,\ 5$ are given in \cite{Bazavov:2017dus}; the coefficients $q_7$ and $s_7$ 
are given in \cite{Bazavov:2020bjn}.
With these constraints we arrive at Taylor series in terms of the baryon chemical potential only,
\begin{equation}
    \chi_n^B(T,\hmu_B) =\sum_{k}\frac{\cb_n^{B,k}(T)}{k!}\hmu_B^k \; ,
\end{equation}
where $n=0$ corresponds to the Taylor series
for the $\hmu_B$-dependent part of the pressure, $n=1$ gives the net baryon-number density and $n>2$ gives higher order cumulants of net baryon-number fluctuations,
\begin{eqnarray}
\chi_0^B(T,\hmu_B)&\equiv&  \frac{P(T,\mu_B)-P(T,0)}{T^4} =
\sum_{k=1}^{\infty} P_{2k}(T) \hmu_B^{2k} \ ,
\label{Pn} \\
\chi_1^B(T,\hmu_B)&\equiv& \frac{n_B(T,\mu_B)}{T^3} =
\sum_{k=1}^{\infty} N_{2k-1}^B(T) \hmu_B^{2k-1}\;\; , \label{nXneutral}\\
\chi_2^B(T,\mu_B) &=& \sum_{k=0}^{\infty} \tilde{\chi}_2^{B,k}(T) \hmu_B^{2k} 
\;\; ,
\label{chineutral}
\end{eqnarray}
with $P_{2k}\equiv \cb_0^{B,2k}/(2k)!$ and
$N_{2k-1}^B=\cb_1^{B,2k-1}/(2k-1)!$. 
The expansion coefficients $\cb_n^{B,k}$ are simply related to the expansion coefficients $\tilde{\chi}_n^{B,k}$ defined in \cite{Bazavov:2020bjn},
\begin{equation}
    \tilde{\chi}_n^{B,k} \equiv \frac{{\cb}_n^{B,k}}{k!} \; .
\end{equation}
For convenience we us here ${\cb}_n^{B,k}$ rather than $\tilde{\chi}_n^{B,k }$ as this emphasizes the 
close relation of the constraint expansion coefficients to the standard cumulants of net baryon-number fluctuations $\chi_k^B$ which equal
${\cb}_n^{B,k}$ in the case $\mu_Q=\mu_S=0$. 
Explicit expression for $\bar{\chi}_n^{B,k}$ are 
given in Appendix A of \cite{Bazavov:2020bjn} for
$k\le 7$. For $k=8$ we give the expansion coefficient ${\cb}_n^{B,8}$ here in Appendix~\ref{app:Pn}.
We also note that in the case $\mu_Q=\mu_S=0$ as well as in the isospin symmetric case, $r=1/2$, the expansion coefficients for the pressure and number density series are closely related, 
\begin{equation}
    N_{2k-1}^B(T)=2k P_{2k}(T) \; .
\end{equation}
In fact, in the case 
$\mu_Q=\mu_S=0$ the expansion coefficients of all higher order 
cumulants are simply related to those of the pressure series, $\cb_n^{B,k}=\dfrac{(k+n)!}{k!}\cb_0^{B,k+n}$. The expansion coefficients shown
in Fig.~\ref{fig:taylor8} thus are 
sufficient to construct the expansions
for $P/T^4$ ($n=0$), $n_B/T^3$ ($n=1$)
and $\chi_2^B$ ($n=2$).
In the strangeness neutral case,
$\mu_Q=0,n_S=0$, the above relation 
only holds for $n=1$. We thus still
need to give results for the 
expansion coefficients of $\cb_2^{B,k}$
with $k=2,\ 4$, $6$. We show these 
expansion coefficients in Appendix~\ref{app:chi2B}. As expected, the qualitative features of the temperature dependence of $\cb_2^{B,k}$
in the $n_S=0$  and $\mu_S=0$ cases 
are similar, {\it i.e.} they behave like
$\chi_{k+2}^B$.

In Fig.~\ref{fig:taylor8} we show results for $\cb_0^{B,2k}$ for the two different cases considered throughout this paper, {\it i.e.} we work in the isospin symmetric case, corresponding to $\mu_Q=0$, and 
consider for the strangeness sector (i) the case
$\mu_S=0$ (left) and (ii) the strangeness neutral case $n_S=0$ (right), respectively. Continuum extrapolated result for the leading order expansion coefficient of the pressure series, $\cb_{0}^{B,2}$, are shown in the two panels on the top of Fig.~\ref{fig:taylor8}. 
They are based on datasets generated on lattices with temporal extent $N_\tau=6,\ 8,\ 12$ and $16$.
Results for the case $\mu_Q=\mu_S=0$ at $T\gsim 135$~MeV had been shown already in Ref.~\cite{Bollweg:2021vqf}
we added here our results at $T=125$~MeV obtained on lattices with temporal extend
$N_\tau=8$, which have not been used in the 
continuum extrapolations. The insets given in these figures for $\chi_2^B$ (left) as well as  $\cb_0^{B,2}$~(right) show  comparisons with the same cumulants calculated in 
a hadron resonance gas (HRG) model
describing the thermodynamics of non-interacting, point-like hadrons\footnote{Throughout this work we use a model based on non-interacting, point-like hadrons listed in the QMHRG2020
list \cite{Bollweg:2021vqf} as HRG model reference system. Such models have been improved by incorporating interactions as described by the S-matrix \cite{Lo:2017ldt} or more phenomenological through the use of finite volumina for baryons \cite{Vovchenko:2016rkn}.}. 
This calculation uses the hadron spectrum compiled in the QMHRG2020 list \cite{Bollweg:2021vqf}. 

For the ${\cal O}(\hmu_B^4)$ expansion coefficients we show in Fig.~\ref{fig:taylor8} continuum extrapolations based on
$N_\tau=6,\ 8$ and $12$ datasets. For the higher
order expansion coefficients we only use results
from our high statistics calculations on lattices
with temporal extent $N_\tau=8$, where more than
1.5 million gauge field configurations\footnote{These datasets have been generated using a Rational Hybrid Monte Carlo Algorithm (RHMC) \cite{Clark:2006fx,MILC:2010pul}. They contain 
gauge field configurations that have been stored after 10 subsequent RHMC time units. 
The actual code package used for our calculations
is described in \cite{Altenkort:2021fqk}.}  have been generated at each temperature value. Results for 
larger $N_\tau$ are consistent with these results but have significantly larger statistical errors. However, as can be seen from the lower order expansion coefficients, cut-off effects are generally small for expansion coefficients at non-zero values of $\hmu_B$. 
The interpolating curves for the 
${\cal O}(\mu^6_B)$ and ${\cal O}(\mu^8_B)$
expansion coefficients shown in Fig.~\ref{fig:taylor8} are cubic spline interpolations.

\subsection{Cumulants and the EoS at non-zero \boldmath$\mu_B$}

From the temperature dependence of the 
${\cal O}(\hmu_B^2)$ expansion coefficient of the pressure shown in Fig.~\ref{fig:taylor8} it is apparent  that deviations from the thermodynamics of 
a non-interacting HRG reach about 20\% at the
pseudo-critical temperature of (2+1)-flavor QCD, $T_{pc}(0)=156.5(1.5)$~MeV \cite{HotQCD:2018pds}
and rapidly become larger at higher temperatures.
Below $T_{pc}$ the leading order expansion coefficient agrees quite well with HRG model calculations
\cite{Bollweg:2021vqf}. As can be seen also in Fig.~\ref{fig:taylor8},
already the ${\cal O}(\hmu_B^4)$ Taylor coefficient deviates from HRG model results more strongly than the ${\cal O}(\hmu_B^2)$ expansion coefficient. For all temperatures in the range
$135~{\rm MeV}\le T \le 165~{\rm MeV}$ the 
sixth and eighth order expansion coefficients are
negative, in contrast to the non-interacting HRG
expansion coefficients, which are all positive.
Even at low temperatures we thus expect to find
that deviations from HRG model calculations 
increase with increasing values of the baryon
chemical potential.

\begin{figure}[t]
\includegraphics[scale=0.6]{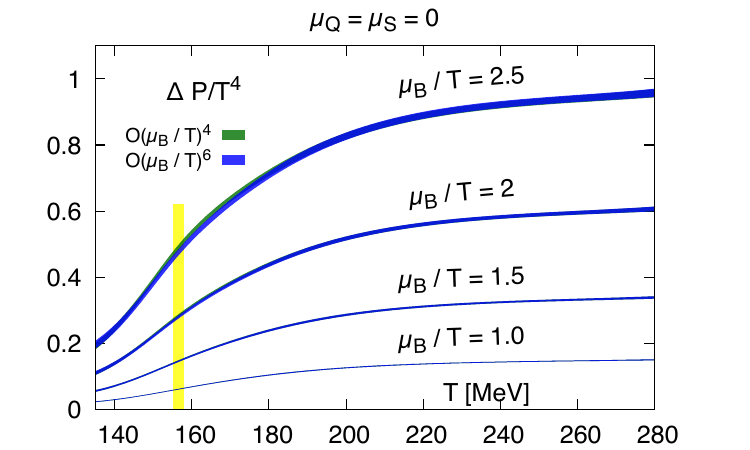}
\includegraphics[scale=0.6]{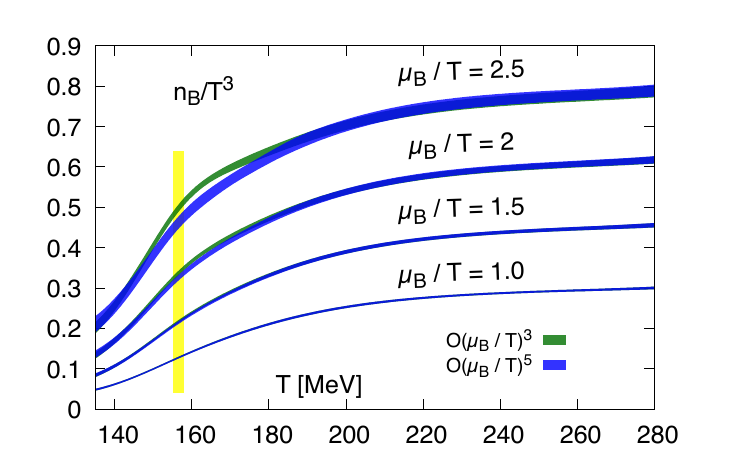}
\caption{Fourth and sixth order Taylor series results for the pressure and corresponding third and fifth order expansion results for the net baryon-number density as function of temperature 
for the case $\mu_Q=\mu_S=0$.}
\label{fig:mudep2}
\end{figure}

Compared to our earlier analysis of the QCD equation of 
state, presented in \cite{Bazavov:2017dus}, the new results 
for the expansion coefficients shown in Fig.~\ref{fig:taylor8}  
are based
on 10 times higher statistics for $N_\tau =8$ and $12$
and include also data on lattices with temporal extent
$N_\tau =16$. This allows to determine also
the contribution from $8^{\rm th}$ order expansion
coefficients to Taylor series of various thermodynamic observables. The highly improved statistics results in a huge improvement of the current calculation over that published 
previously \cite{Bazavov:2017dus}. 
We update in Fig.~\ref{fig:mudep2} our results for the pressure 
and the net
baryon-number density calculated in sixth and fifth order of the Taylor expansion, respectively. Results are shown as function of temperature for the case $\mu_Q=\mu_S=0$ using the 
continuum extrapolated data for $\chi_2^B$ and $\chi_4^B$,
as well as the spline interpolated data for $\chi_6^B$, obtained 
on lattices with temporal extent $N_\tau=8$.
Obviously, the ``wiggly" structure seen in
the old calculations for ${\cal O}(\hmu_B^6)$ expansions at $\hmu_B=2.5$ \cite{Bazavov:2017dus} is smoothed out in our new 
high statistics analysis and the ${\cal O}(\hmu_B^6)$ results 
agree well with ${\cal O}(\hmu_B^4)$ expansions in the entire temperature range.

\begin{figure}[t]
\includegraphics[scale=0.45]{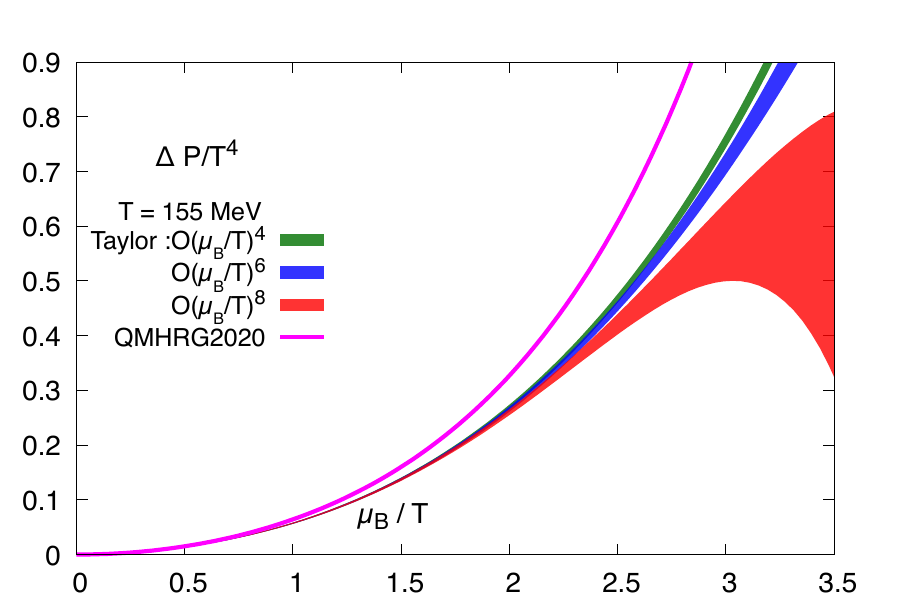}
\includegraphics[scale=0.45]{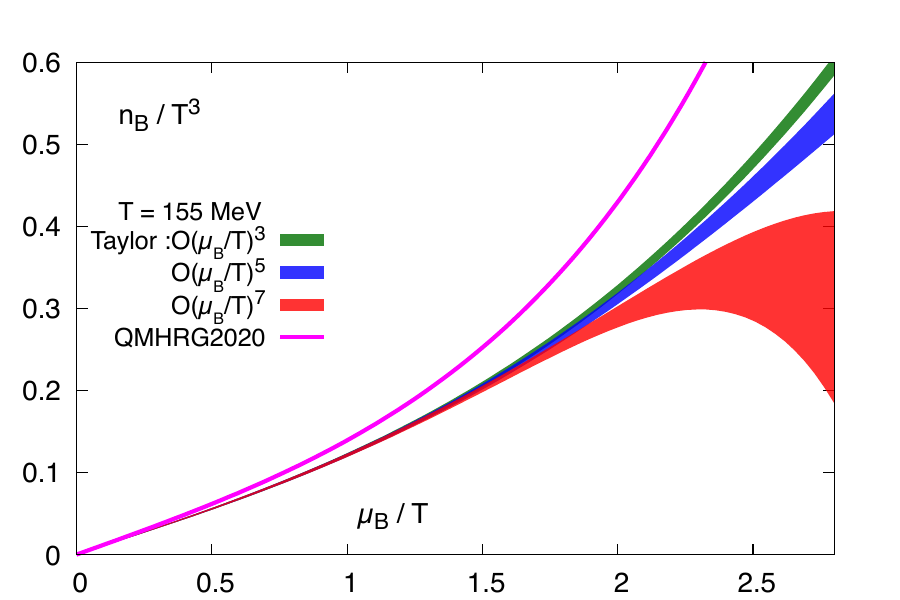}
\includegraphics[scale=0.45]{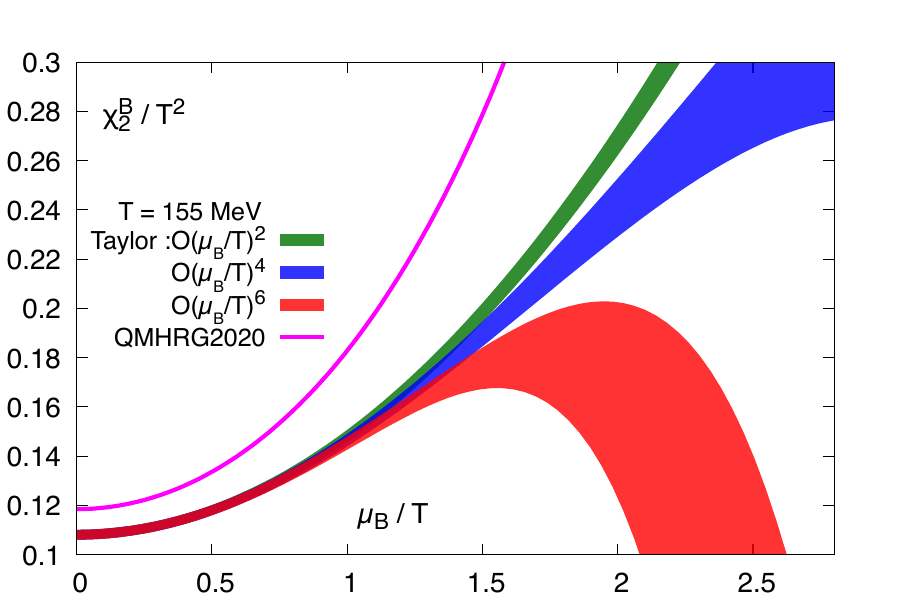}
\caption{The pressure (top), net baryon-number
density (middle) and net baryon number fluctuations (bottom) versus $\hmu_B$ 
at $T=155$~MeV.}
\label{fig:mudep1}
\end{figure}

On the basis of a sixth order Taylor expansion we thus have no
indications for a radius of convergence being smaller than $\hmu_B=2.5$ nor do we have indications for a poor convergence of 
the Taylor expansions of $P/T^4$ and $n_B/T^3$, respectively. 
This will change
when discussing the eighth order contribution to the
Taylor series. We stress, however, already here that we need
to distinguish between the radius of convergence of the Taylor
series, which is the same for all observables determined as 
derivatives of $P/T^4$ with respect to $\hmu_B$, and the rate of convergence of expansions for theses observables to their asymptotic values, which will be slower with increasing order
of the derivatives.

Taking also into account the contribution from eighth order Taylor expansion coefficients of the pressure we show in Fig.~\ref{fig:mudep1} results for 
the $\hmu_B$-dependence of the pressure,
net baryon-number density and the second order cumulant of net baryon-number
fluctuations. Shown are results obtained by using different orders of the 
Taylor expansion at a fixed value
of the temperature in the vicinity of $T_{pc}$, {\it i.e.} $T=155$~MeV, for the case $\mu_Q=\mu_S=0$. As can be seen 
deviations from QMHRG2020 increase with increasing $\hmu_B$ and these deviations are larger for higher order cumulants. It also is apparent from this figure, that the rate of convergence of the expansions for higher order cumulants slows down. Being limited to a certain order in the expansion thus allows to give reliable results
for higher order cumulants only in a smaller 
$\hmu_B$ range, although the expansions for 
all cumulants have the same radius of convergence. We will give a more quantitative
discussion of the $\hmu_B$-range, in which 
the current Taylor expansions for different
cumulants are expected to give reliable results, in Section~\ref{sec:compare}.

\begin{figure*}[t]
\begin{center}
\includegraphics[scale=0.56]{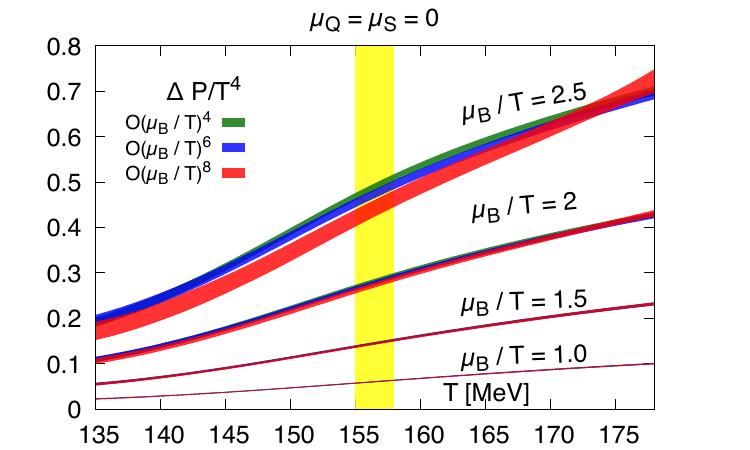}
\includegraphics[scale=0.56]{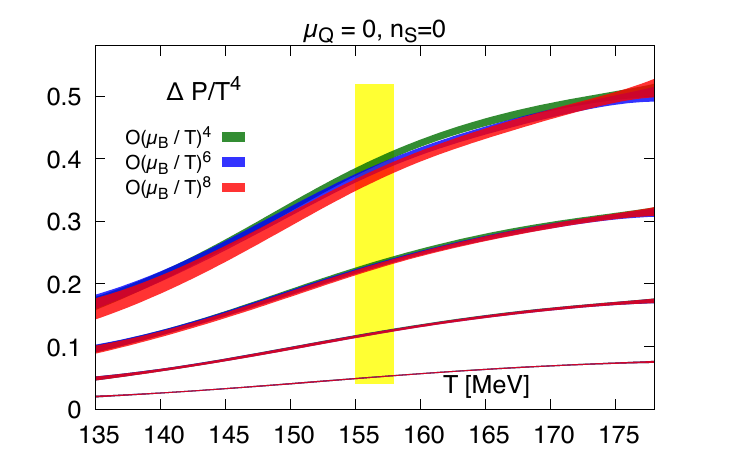} 
\includegraphics[scale=0.56]{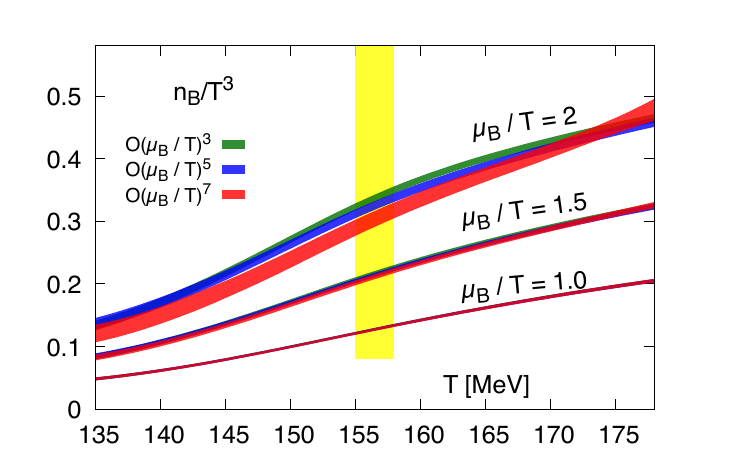}
\includegraphics[scale=0.56]{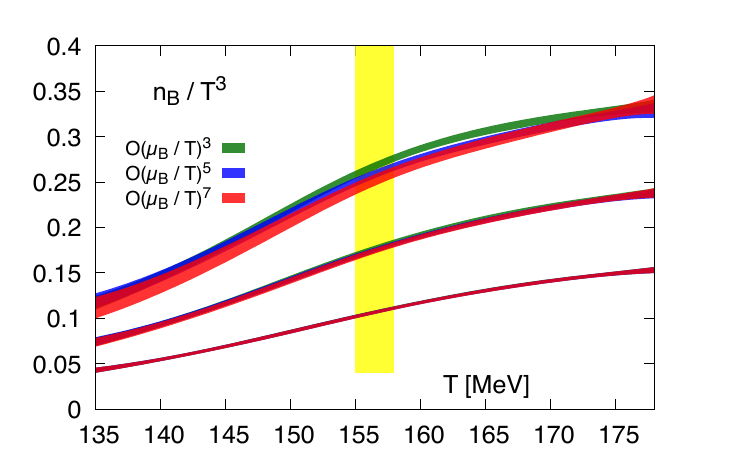}
\end{center}
\caption{Pressure (top) and net baryon-number density (bottom) versus temperature
for several values of the baryon chemical potential. Figures on the left correspond to the case $\mu_Q=\mu_S=0$ and on the right to
the strangeness neutral, isospin symmetric case.
The Taylor expansions are based on the continuum and spline interpolated data shown in Fig.~\ref{fig:taylor8}.
}
\label{fig:EoS}
\end{figure*}

In Fig~\ref{fig:EoS} we show the $\hmu_B$-dependent contribution
to the pressure as function of
temperature for some values of $\hmu_B$, {\it i.e.} for $\hmu_B= 1.0,\ 1.5,\ 2.0$ and $2.5$, and for the  net baryon-number density
for $\hmu_B= 1.0,\ 1.5,\ 2.0$.
In all cases we show results
obtained in different orders of the Taylor series expansion. 
For these values of the baryon chemical potential the 
${\cal O}(\hmu_B^8)$ Taylor series for the pressure agrees well with the lower
order results. For $n_B/T^3$ we do not show results 
from an ${\cal O}(\hmu_B^8)$ at $\hmu_B=2.5$ as it is apparent 
from Fig.~\ref{fig:mudep1} that higher order expansion
coefficients will be needed to obtain reliable 
results for the pressure at that large value of $\hmu_B$.

In the entire temperature range analyzed by us
the Taylor series for pressure converges well for 
$0\le \hmu_B\le 2.5$. For the number density this can be 
stated at present only for the range $0\le \hmu_B\le 2.0$.
Although that may turn out to be somewhat larger once the 
statistical accuracy in calculations of eighth order expansion
coefficients is increased. Nonetheless,
based on the analysis of $8^{\rm th}$ order Taylor
expansions of the pressure we have no hint for a
radius of convergence
smaller than $\hmu_B\ \sim 2.5$ that would limit the
applicability of Taylor expansions at temperatures 
$T\gsim 125$~MeV.

\section{Radius of convergence and Pad\'e approximations}
\label{radius}

In general the radius of convergence of the Taylor series for a function
\begin{equation}
f(x)= \sum_n^{\infty} c_n x^n
\label{series}
\end{equation}
is given by the location of a singularity of $f$ in the complex 
$x$-plane that is closest to the origin. Of course, 
rigorous statements on the radius of convergence 
of a Taylor series can only be made by analyzing the asymptotic behavior of the expansion coefficients in the limit $n\rightarrow \infty$. 
Having at hand only a few expansion coefficients of the Taylor series for the pressure in QCD we naturally can only obtain estimators for the radius of convergence and extract some information on the analytic structure of thermodynamic functions in the plane of complex chemical potentials.

We are dealing with Taylor series in terms of $\hmu_B$ for which only every
second expansion coefficient is non-zero,
{\it e.g.} the pressure series which has 
non-vanishing expansion coefficients $\cb_0^{B,n}$ only for even $n$.
The simplest estimator, $r_{c,n}$, for the radius of convergence, 
$r_c=\lim\limits_{n\rightarrow \infty} r_{c,n}$, is obtained from subsequent, 
non-vanishing expansion coefficients. We define $r_{c,n}=\sqrt{|A_n|}$, with
\begin{equation}
 A_n= \frac{c_{n}}{c_{n+2}} \;\; , \;\; n\;{\rm even} \; .
\end{equation}
Another frequently used estimator, with improved convergence 
properties, has been introduced by Mercer and Roberts \cite{Mercer:1990}, $r_{c,n}^{MR} = |A_n^{MR}|^{1/4}$, with
\begin{equation}
A_n^{MR}= s\frac{c_{n+2} \, c_{n-2} - c_n^2 } { c_{n+4} \, c_n - c_{n+2}^2 } \;\; , \;\; n\; {\rm even} \; .
\label{rcMR}
\end{equation}
The estimators based 
on the ratios $A_n$ and $A_n^{MR}$ are
related to poles of [n,2] and [n,4]
Pad\'e approximants for the series
expansion of $f(x)$. We thus will consider the structure of such Pad\'e approximants in the following.

When constructing Pad\'e approximants
for the pressure series of (2+1)-flavor QCD we take advantage of the fact that 
the two leading expansion coefficients 
of the pressure, $P_{2k}=\cb_0^{B,2k}/(2k)!,\ k=1,2$, are 
strictly positive. We thus rescale the 
pressure by a factor $P_4/P_2^2$ and redefine the expansion parameter as
\begin{equation}
    \xb=\sqrt{\frac{P_{4}}{P_2}}\ \hmu_B\equiv \sqrt{\frac{\cb_0^{B,4}}{12\cb_0^{B,2}}}\ \hmu_B \; .
\end{equation}
This allows us to re-write the expansion of the 
pressure in terms of expansion coefficients 
\begin{equation}
 c_{2k,2}=\frac{P_{2k}}{P_2} \left(\frac{P_2}{P_{4}}\right)^{k-1}\; ,   \label{c2k2}
\end{equation}
which gives $c_{2,2}=c_{4,2}=1$.  Therefore, for  $\mu_Q=\mu_S=0$ as well as for the strangeness neutral case, the analytic structure of the QCD pressure, that one can deduce from an eighth order Taylor series in QCD, entirely depends on two expansion parameters, 
\begin{eqnarray}
    c_{6,2}&=&\frac{P_6 P_2}{P_4^2} = \frac{2}{5} \frac{\cb_6^B \cb_2^B}{(\cb_4^B)^2} \;\; ,\\
    c_{8,2}&=& \frac{P_8P_2^2}{P_4^3} = \frac{3}{35} \frac{\cb_8^B (\cb_2^B)^2}{(\cb_4^B)^3}
    \;\; .
    \label{parameter}
\end{eqnarray}
With this we obtain
\begin{eqnarray}
\frac{(\Delta P(T,\mu_B)/T^4) P_4}{P_2^2} &=&
\sum_{k=1}^{\infty} c_{2k,2} \xb^{2k}\; ,
\\
&=& \xb^2+\xb^4+ c_{6,2} \xb^6
+ c_{8,2} \xb^8 + ... \nonumber
\;   \; ,
\end{eqnarray}
with $\Delta P(T,\mu_B)=P(T,\mu_B)-P(T,0)$.

The two diagonal Pad\'e approximants, that can be constructed from our eighth order series
for the pressure are given by,
\begin{eqnarray}
    \hspace*{-0.5cm}P[2,2] &=& \frac{\xb^2}{1-\xb^2}\;\; ,  \label{pd22}\\
    \hspace*{-0.5cm}P[4,4] &=&  \frac{(1-c_{6,2}) \xb^2 +
\left(1 - 2 c_{6,2} +c_{8,2} \right) \xb^4}{(1-c_{6,2})+
(c_{8,2}-c_{6,2}) \xb^2 + (c_{6,2}^2 - c_{8,2}) \xb^4} .
\label{pd44}
\end{eqnarray}

The [2,2] Pad\'e has a pole on the real axis for
$\xb^2=1$, {\it i.e.} for
$\mu_{B,c}\equiv r_{c,2}=\sqrt{12\cb_2^B/\cb_4^B}$ which is the standard ratio estimator for the radius of 
convergence. The [4,4] Pad\'e has four poles which come in two pairs, corresponding to zeroes of the polynomial in the denominator of Eq.~\ref{pd44} which is quadratic in $z\equiv \xb^2$. The two zeroes in $z$ are given by 

\begin{figure*}[t]
	\begin{center}
\includegraphics[scale=0.36]{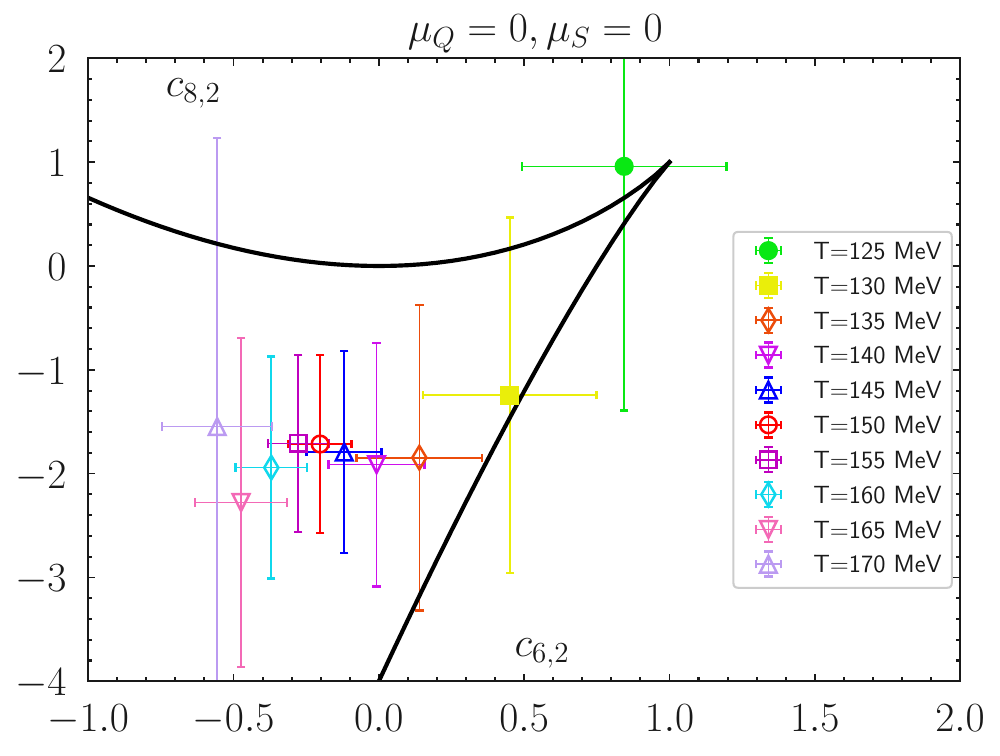}
\includegraphics[scale=0.36]{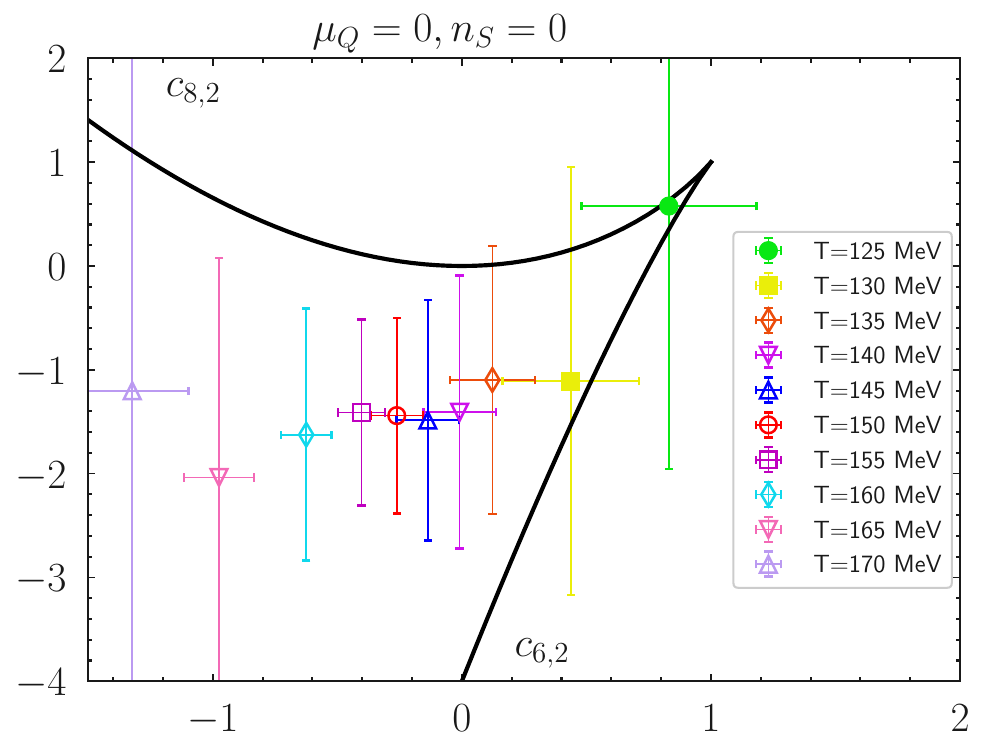}
\end{center}
\caption{The expansion coefficients $c_{8,2}$ versus $c_{6,2}$ on lattices with temporal extent $N_\tau=8$
in the entire temperature range $125~{\rm MeV}< T < 175~{\rm MeV}$ covered in our calculations. The area
bounded by the two black lines indicates the region in parameter space, in which all poles of the [4,4] Pad\'e are complex. The left hand figure corresponds to the case $\mu_Q=\mu_S=0$ and the right hand figure is for the strangeness neutral, isospin symmetric case.}
\label{fig:c8c6plane}
\end{figure*}	

\begin{eqnarray}
z^\pm &=&\frac{c_{8,2} - c_{6,2} \pm \sqrt{
(c_{8,2}-c_{8,2}^+) (c_{8,2}-c_{8,2}^-)}}{2 (c_{8,2} -c_{6,2}^2)} \;\; ,
\label{zeroes}
\end{eqnarray}
with
\begin{equation}
c_{8,2}^\pm = -2 + 3 c_{6,2} \pm 2(1-c_{6,2})^{3/2}
\;\; .
\label{c8pm}
\end{equation}
It is easy to see that the argument of the square root appearing in Eq.~\ref{zeroes} is positive for $c_{6,2}>1$. Complex zeroes with ${\rm Re}(\hmu_B)\ne 0$ thus exist only for
\begin{equation}
(i)\;\;	c_{6,2}<1 \;\; {\rm and}\;\;  c_{8,2}^{ -} < c_{8,2} < c_{8,2}^+ \;.
	\label{czeroes}
\end{equation}
Outside this region the zeroes $z^\pm$ are real and thus correspond to pairs of real poles in terms of $\hmu_B$ if $z^\pm>0$ and purely imaginary poles if $z^\pm<0$. In fact, as we show in Appendix~\ref{app:zpzm}, there is a  small region in parameter space
$(c_{6,2},c_{8,2})$, close to $c_{8,2}^+$ in which $z^+<0$
and $z^-<0$, 
\begin{equation}
(ii)\;\;    c_{6,2}<0\;\; {\rm and}\;\; 
    c_{8,2}^{+} < c_{8,2} < c_{6,2}^2
    \; .
\end{equation}
This leads to 
two pairs of purely imaginary poles in $\hmu_B$. Everywhere else in parameter space at least one pair of real zero exists, which, however, not always is the
pair of zeroes closest to the origin (see Appendix~\ref{app:zpzm}).
	
In order to get further information on the poles 
of the [4,4] Pad\'e approximant for the pressure
and, in particular, deduce conditions for the occurrence of real poles we show in Fig.~\ref{fig:c8c6plane} results for $c_{8,2}$ versus $c_{6,2}$ obtained in the 
temperature range\footnote{We do not show results for $T=175$~MeV as errors are even larger than those shown for $T=170$~MeV.} $125~{\rm MeV}\le T\le 170~{\rm MeV}$ from the spline interpolated $N_\tau=8$ expansion coefficients, $\cb_0^{B,6}$, $\cb_0^{B,8}$, and the continuum extrapolations for $\cb_0^{B,2}$, $\cb_0^{B,4}$ shown in Fig.~\ref{fig:taylor8}. Also shown in this figure are the boundaries for the triangular shaped region, bounded by $c_{8,2}^+$ and $c_{8,2}^-$,
inside which only complex poles exist for the [4,4] Pad\'e of the eighth order Taylor series of the pressure.
We show results for the case
$\mu_Q=\mu_S=0$ (left) and  $\muQ=0$, $n_S=0$  (right), respectively. 	
	
As can be seen in Fig.~\ref{fig:c8c6plane},  despite the currently large errors on the location of the poles, it is well established  that the poles occur in
the complex $\hmu_B$-plane for all temperatures $135~{\rm MeV}\le T\le 165~{\rm MeV}$. 
Within our current statistical errors we cannot rule out that
pairs of real and/or purely imaginary poles will occur at
temperatures below $T=135$~MeV as well as for temperatures above $T=165$~MeV. 
In fact, this is expected to be the case at low enough temperatures, where one can see in Fig.~\ref{fig:taylor8} that $\cb_0^{B,6}$ and 
$\cb_0^{B,8}$ become positive at $T\simeq 125$~MeV, and also at high
temperature where Fig.~\ref{fig:taylor8} shows that $\cb_0^{B,6} < 0$ while $\cb_0^{B,8}>0$ at $T\simeq 175$~MeV.

\begin{figure*}[t]
	\begin{center}
\includegraphics[scale=0.42]{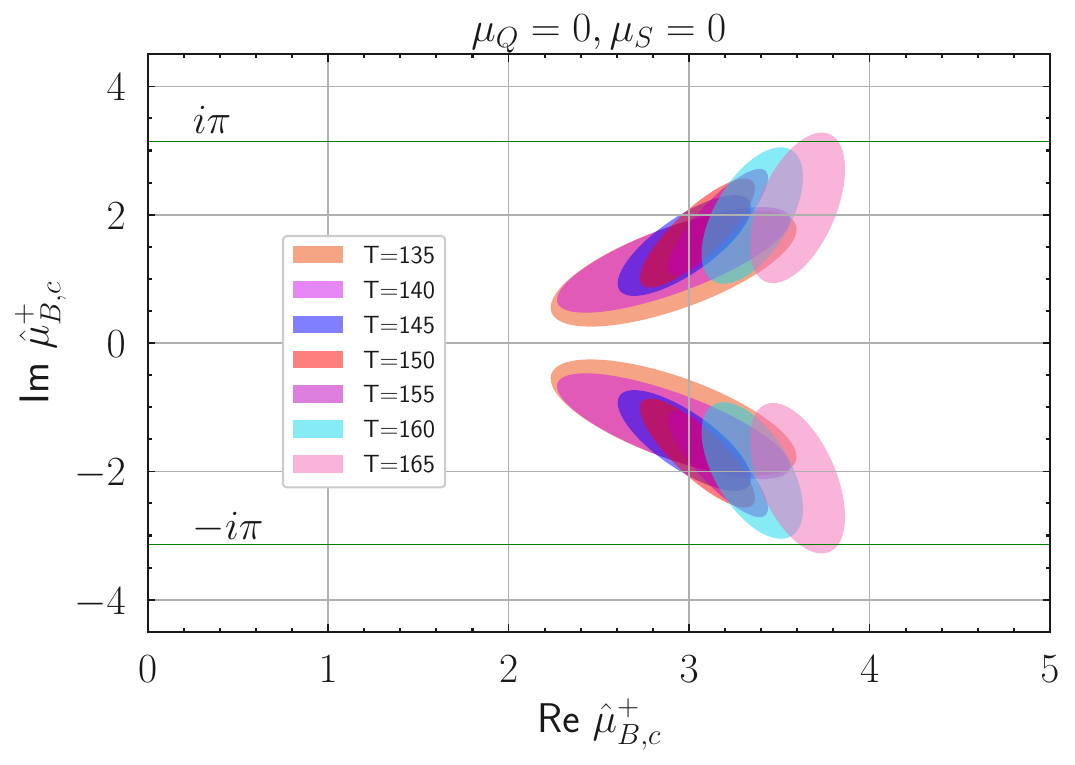}
	\includegraphics[scale=0.42]{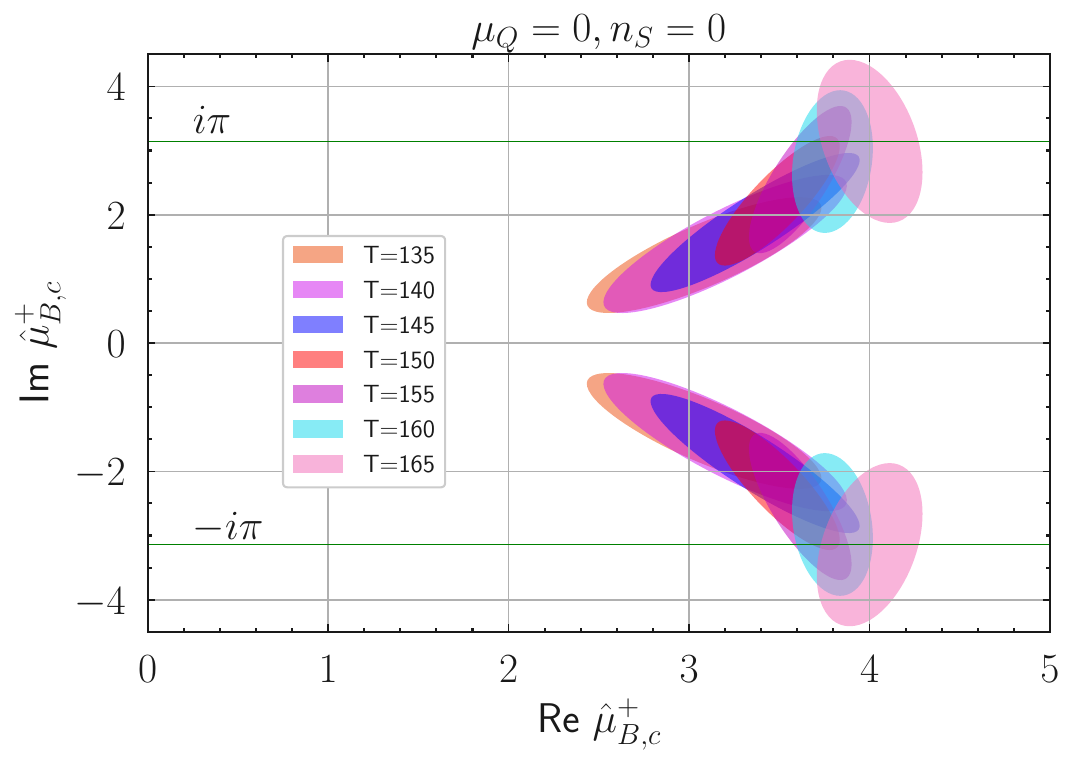}

\end{center}
\caption{Location of poles nearest to the origin obtained from the [4,4] Pad\'e approximants  in the complex $\hmu_B$-plane. Only poles with ${\rm Re} (\mu_B)>0$ are shown. Shown are results the case $\mu_Q=\mu_S=0$ (left) and the strangeness neutral, isospin symmetric case (right).
}
\label{fig:mupoles}
\end{figure*}

At low temperatures the complex valued poles 
leave the area bounded by $c^\pm_{8,2}$ in a region of parameter space where $c_{6,2}>0$ or correspondingly $\cb_0^{B,6}>0$.
As discussed and also indicated in the figure shown in  Appendix~\ref{app:zpzm} for $c_{6,2}>0$ there is a small region in parameter space above 
the boundary defined by $c_{8,2}^+$ where all poles are 
strictly real, before entering the region where a 
pair of real and imaginary poles exists. One can check
that for all $0<c_{6,2}<1$ the real pole actually is closer to the origin as long as $c^2_{6,2}<c_{8,2}<c_{6,2}$. Our results on the temperature dependence of $c^2_{6,2}$ and $c^2_{8,2}$  thus suggest that with decreasing temperature a pair of complex poles moves towards the real axis and gives rise to two real poles. With decreasing temperature one of these real poles moves towards infinity and comes back as an imaginary pole with
large magnitude.

In the region where the conditions given in Eq.~\ref{czeroes} hold
one has four zeroes in terms of $\xb$ corresponding to the positive and negative roots of $z^\pm$. They yield four poles of the [4,4] Pad\'e in the complex $\mu_B$-plane with non-vanishing
imaginary part of $\hmu_B$. We represent these poles in polar coordinates,
\begin{equation}
    \hmu_{B,c}^\pm = \pm r_{c,4} e^{\pm i \Theta_{c,4}} \; .
    \label{muBplane}
\end{equation}
For temperatures $135~{\rm MeV}\le T\le 165~{\rm MeV}$ the zeroes $z^\pm$ are complex conjugate to each other. In the $\xb$-plane the absolute value of
the distance of the poles from the origin is then given by,
\begin{eqnarray}
 |z^+ z^-|^{1/4}
    &=&\left| 
    \frac{1-c_{6,2}}{c_{6,2}^2-c_{8,2}}\right|^{1/4} \; ,
	\label{estimator}
\end{eqnarray}
which is the Mercer-Roberts estimator, introduced in Eq.~\ref{rcMR}, for a series in the rescaled expansion parameter $\xb$. We note that this  relation between the Mercer-Roberts estimator and the magnitude of $|z^\pm|$ does not hold for 
the case of purely real or purely imaginary poles of the [4,4] Pad\'e
(see discussion in Appendix~\ref{app:zpzm}).
In these cases the 
distances to the origin $|z^+|$ and $|z^-|$ differ
from each other.

\begin{figure}[t]
	\begin{center}
\includegraphics[scale=0.72]{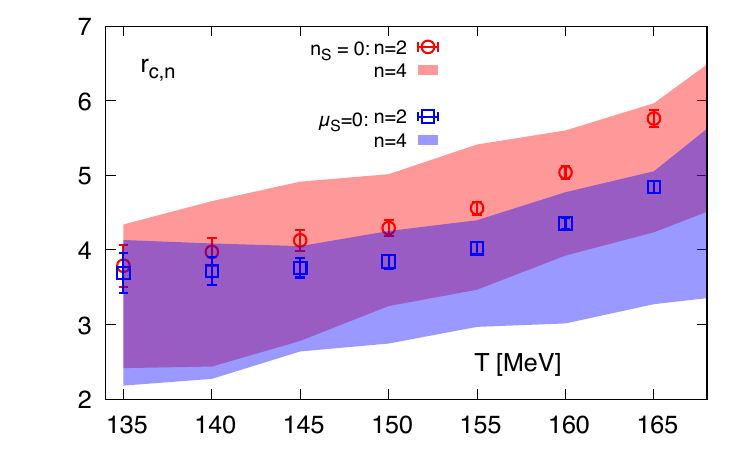}
\end{center}
\caption{Magnitude of poles
nearest to the origin obtained from the [2,2] (squares and circles) and [4,4] (bands) Pad\'e approximants for
Taylor expansions at $\mu_Q=\mu_S=0$ and for strangeness neutral, isospin symmetric media, respectively.}
\label{fig:dist}
\end{figure}

Using Eqs.~\ref{muBplane} and \ref{estimator} we obtain for $c_{6,2} < 1$ the location of the poles in the
complex $\mu_B$-plane
\begin{eqnarray}
    r_{c,4} &=& r_{c,2}\ |z^+ z^-|^{1/4} 
    = \sqrt{\frac{12 \cb_0^{B,2}}{\cb_0^{B,4}}}  
    \left| 
    \frac{1-c_{6,2}}{c_{6,2}^2-c_{8,2}}\right|^{1/4}  \; ,\label{muBplane2} 
    \end{eqnarray}
    \begin{eqnarray}
    \Theta_{c,4} &=& \arccos\left(\frac{c_{6,2}-c_{8,2}}{2\sqrt{(1-c_{6,2})(c_{6,2}^2-c_{8,2})}}\right) \nonumber \\
    &=& \arccos\left(\frac{(c_{6,2}-c_{8,2})\cb_0^{B,4} }{24 (1-c_{6,2}) \cb_0^{B,2}}\ r_{c,4}^2 \right)
    \; .
    \label{muBtheta}
\end{eqnarray}

Expressing the relation given in Eq.~\ref{muBplane2}
in terms of the cumulants $\cb_0^{B,n}$ entering the Taylor series 
for the pressure, Eq.~\ref{Pn}, we have in the region of complex poles,
\begin{eqnarray}
 r_{c,4}	
&=&\left( \frac{8!}{4!}\right)^{1/4} 
	\left| \frac{30(\cb_0^{B,4})^2-12\cb_0^{B,6}\cb_0^{B,2}}{56(\cb_0^{B,6})^2-30\cb_0^{B,8}\cb_0^{B,4}} \right|^{1/4} \; .
\label{rcMR8}
	 \end{eqnarray}

\begin{figure*}[ht]
\includegraphics[scale=0.35]{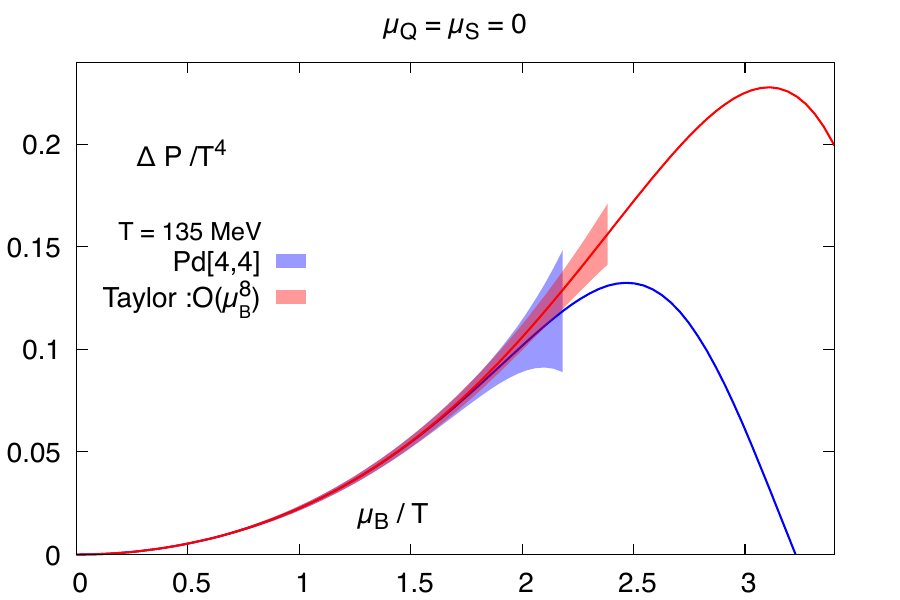}
\includegraphics[scale=0.35]{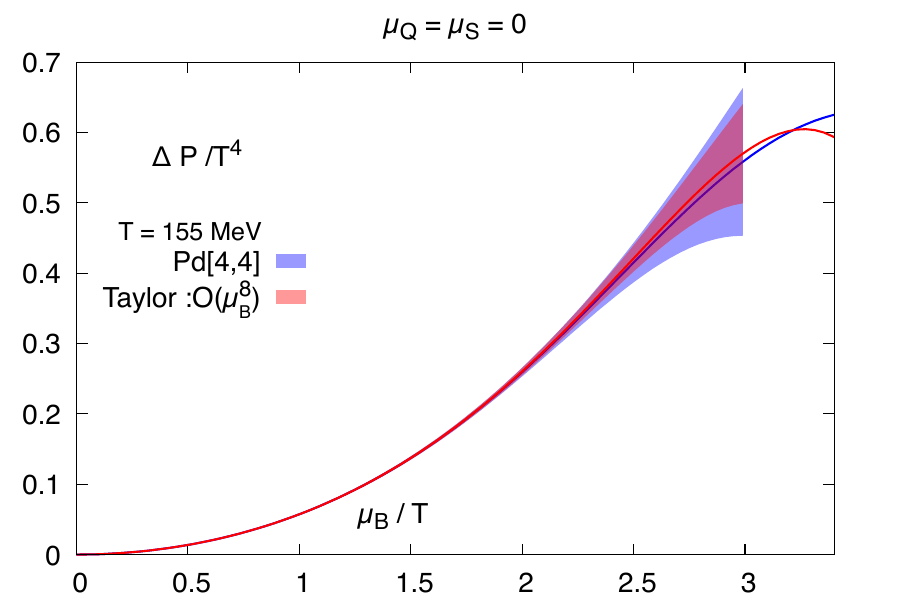}
\includegraphics[scale=0.35]{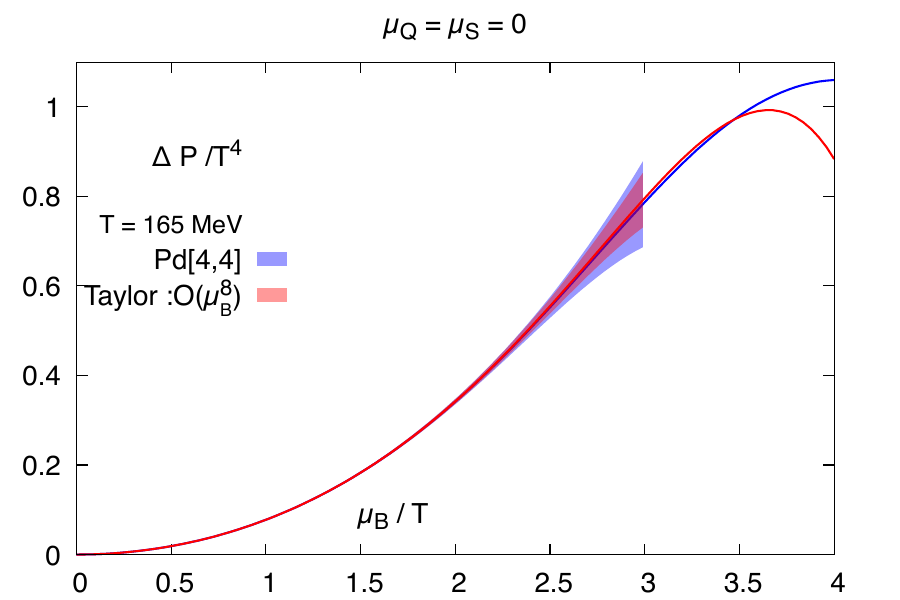}
\includegraphics[scale=0.35]{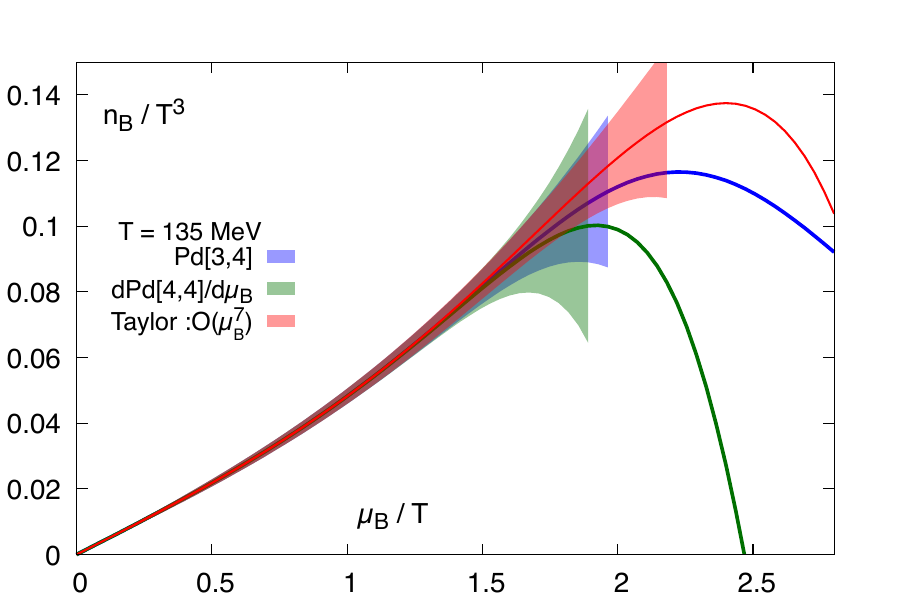}
\includegraphics[scale=0.35]{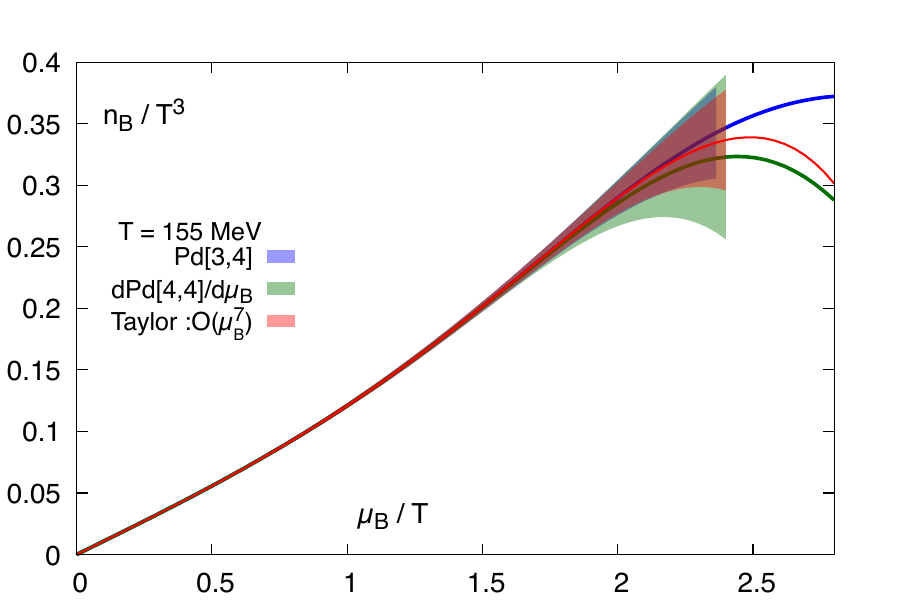}
\includegraphics[scale=0.35]{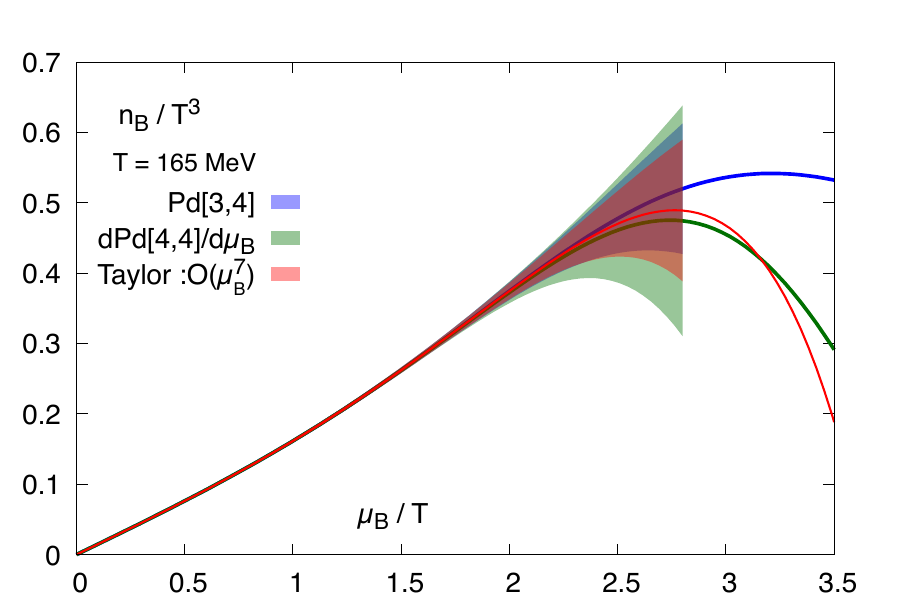}
\includegraphics[scale=0.35]{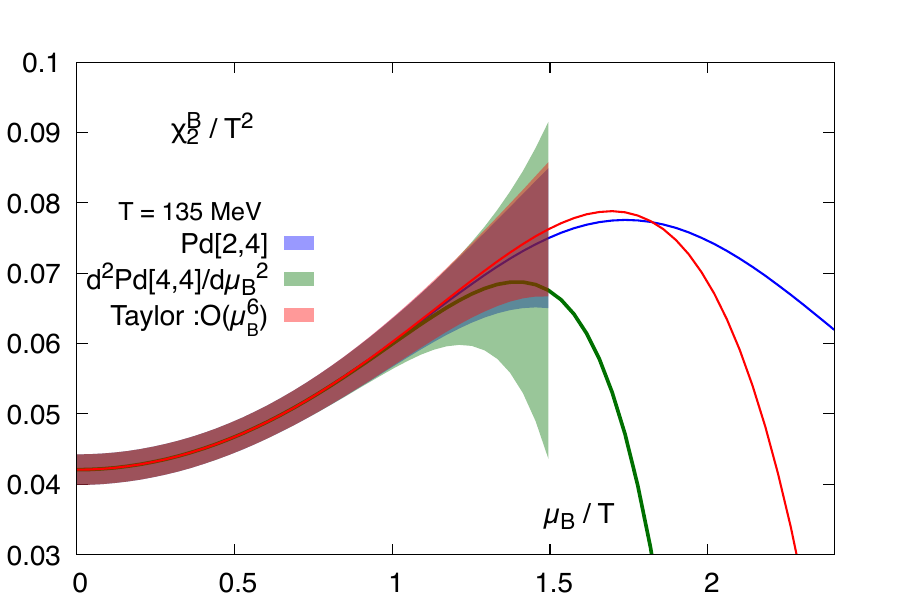}
\includegraphics[scale=0.35]{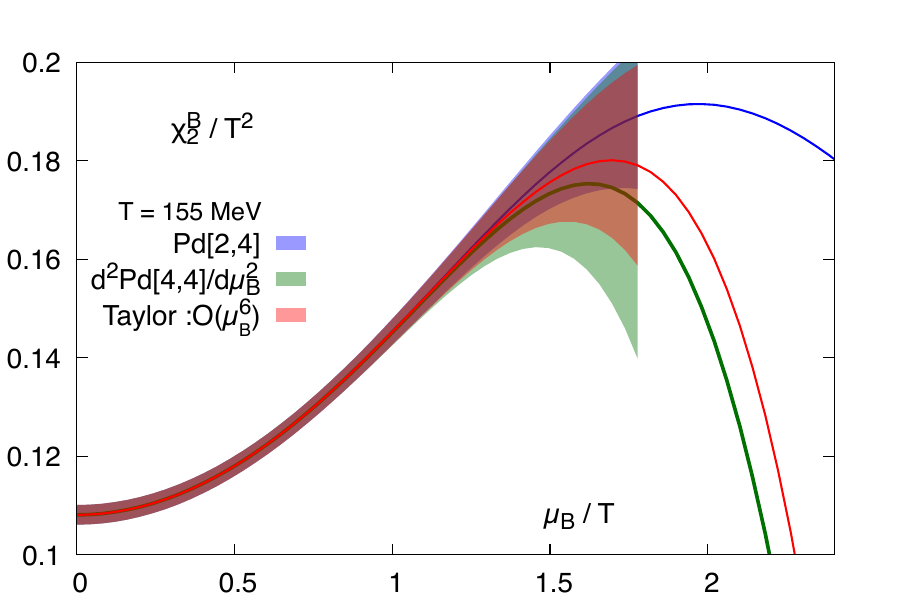}
\includegraphics[scale=0.35]{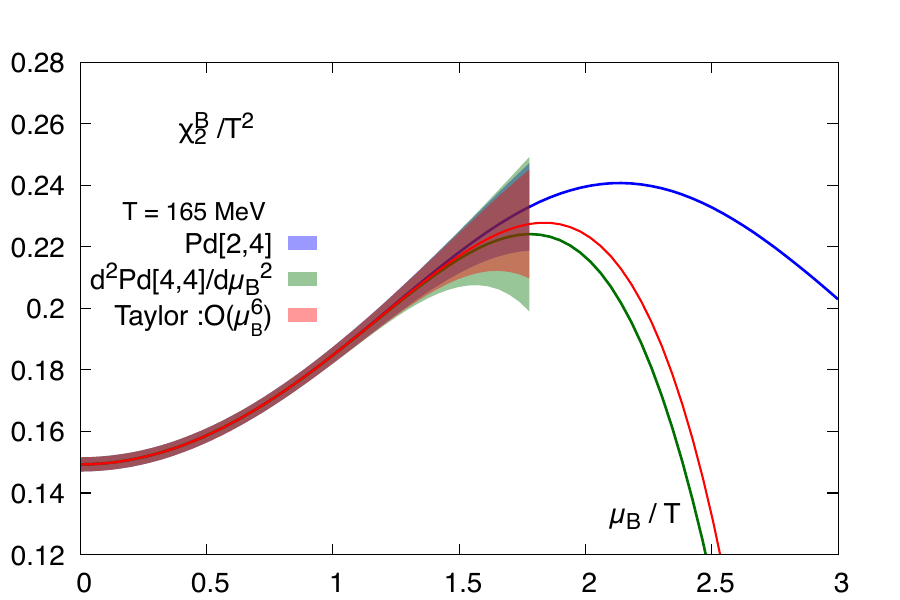}

\caption{Comparison of the [n,4] Pad\'e approximants
for the pressure ($n=4$), the net baryon-number density ($n=3$) and the second order cumulant of net baryon-number fluctuations ($n=2$) with corresponding
Taylor expansions. Shown are results for $T=135$~MeV
(left), $155$~MeV(middle) and  $165$~MeV (right) versus $\hmu_B$ for the case $\mu_Q=\mu_S=0$.
Also shown are derivatives of the [4,4] Pad\'e
approximants with respect to $\hmu_B$ (green bands).
}
\label{fig:padetaylor8}
\end{figure*}

The positions of the poles in the complex $\hmu_B$-plane are shown in Fig.~\ref{fig:mupoles}. Only the two poles in the region ${\rm Re}(\hmu_B)\ge 0$ are shown. With decreasing temperature the poles 
move closer to the real axis as $c_{8,2}$ approaches
$c_{8,2}^+$, {\it i.e.} $\Theta_{c,4}=0$ for $c_{8,2}=c_{8,2}^+$. Furthermore, it is clear from 
Eq.~\ref{muBtheta} that $\Theta_{c,4}$ and $r_{c,4}$
are correlated, which leads to the orientation of the
1-$\sigma$ error ellipse in the complex $\mu_{B,c}$
plane arising from the errors on $c_{6,2}$ and $c_{8,2}$, which are assumed to given by independent Gaussian distributions of the variables
$c_{6,2}$ and $c_{8,2}$.

In Fig.~\ref{fig:dist} we show as symbols and bands, respectively, the distance of poles of the [2,2] and [4,4] Pad\'e approximants
from the origin as function of temperature. 
The bands shown in Fig.~\ref{fig:dist} have been obtained by using the spline interpolations of $\cb_0^{B,6}$ and $\cb_0^{B,8}$ on $N_\tau=8$ lattices and the
continuum extrapolated results for $\cb_0^{B,2}$ and $\cb_0^{B,4}$,
shown in Fig.~\ref{fig:taylor8}, respectively.
As can be seen the two estimators yield a similar magnitude for $r_{c,2}$ and $r_{c,4}$.
Their location in the complex $\mu_B$-plane, however, 
is quite different. While the poles of the [2,2]
Pad\'e are always on the real axis, the poles of the
[4,4] Pad\'e are in the complex plane in the entire
interval $135~{\rm MeV}\le T\le 165~{\rm MeV}$.

For $135~{\rm MeV}\le T\le  165~{\rm MeV}$ we find that the poles of the [4,4] Pad\'e appear at a distance from the origin corresponding to $|\hmu_B| \gsim 2.5$ at $T\simeq 135$~MeV and rises to values larger than $|\hmu_B| \gsim 3$ for $T\gsim T_{pc}$. 
This also are the best estimates for a temperature dependent bound on the radius of convergence of the 
Taylor series for the pressure,
based on the Mercer-Roberts estimator. The information
extracted from the [4,4] Pad\'e approximants on the 
location of poles in the analytic function representing the pressure as function of a complex valued chemical potential $\hmu_B$ thus seems to be consistent with the good convergence properties of the Taylor series itself.

\section{Comparison of Pad\'e approximants and Taylor series}
\label{sec:compare}
In Fig.~\ref{fig:padetaylor8} we compare the [n,4]
Pad\'e approximants for the pressure ($n=4$), the net baryon-number density ($n=3$) and the second order cumulant of net baryon-number fluctuations ($n=2$) with corresponding Taylor expansions that use expansion
coefficients
$\cb_{4-n}^{B,k}$ with $k\le 8$. We show results
obtained at three temperatures in the interval in which 
our results clearly yield complex valued
poles only, {\it i.e.} $T=135, 155, \text{and} \ 165$~MeV,
respectively. As error bands quickly become large for large $\hmu_B$ we show errors only up to the point where relative errors are less then 15\%. In this range of $\hmu_B$ values also the Pad\'e approximants and the straightforward Taylor expansions agree quite well. 
 
\begin{figure*}[t]
\begin{center}
\includegraphics[scale=0.59]{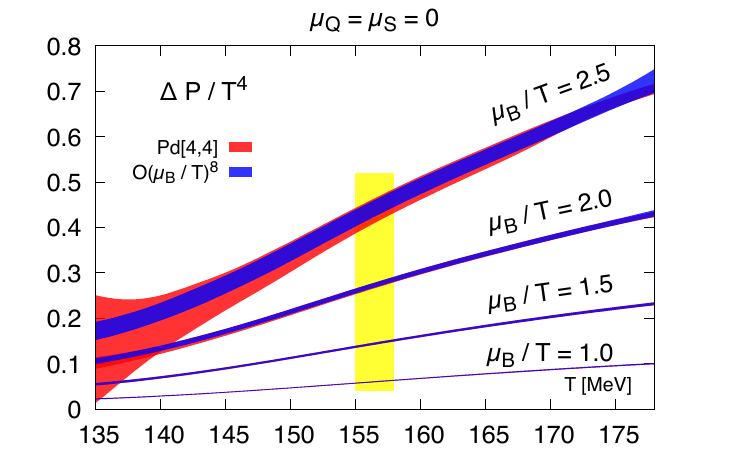}
\includegraphics[scale=0.59]{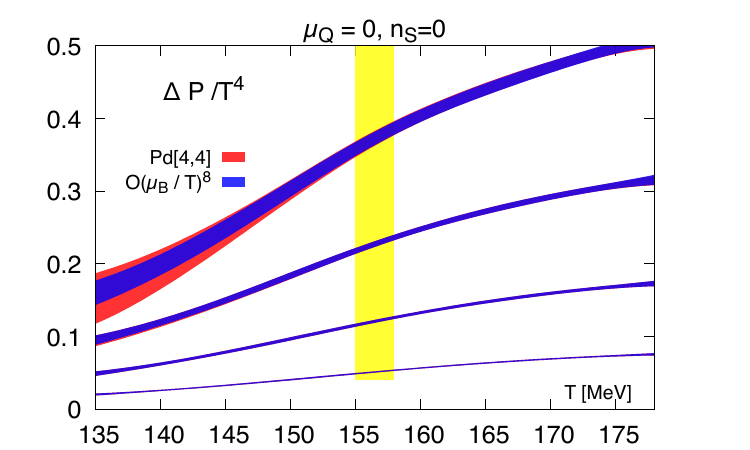}
\includegraphics[scale=0.59]{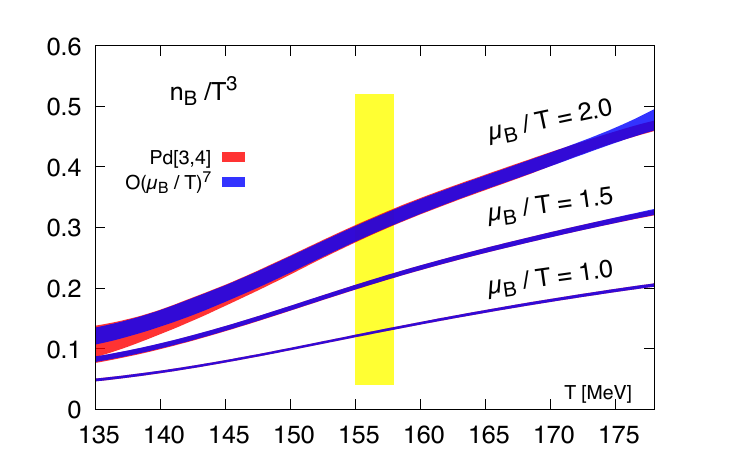}
\includegraphics[scale=0.59]{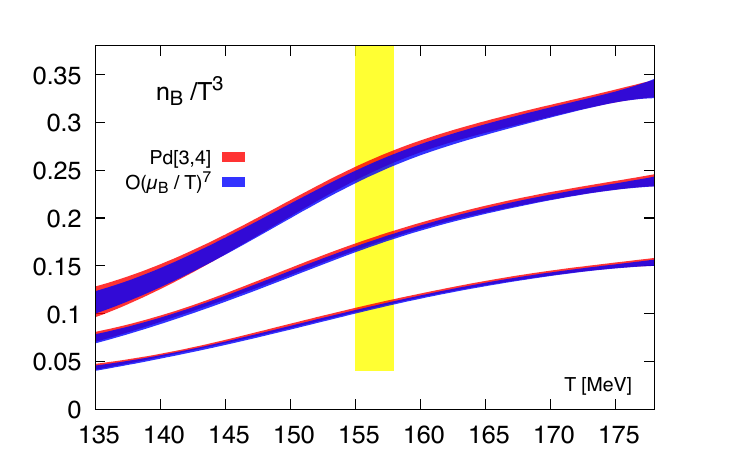}
\includegraphics[scale=0.59]{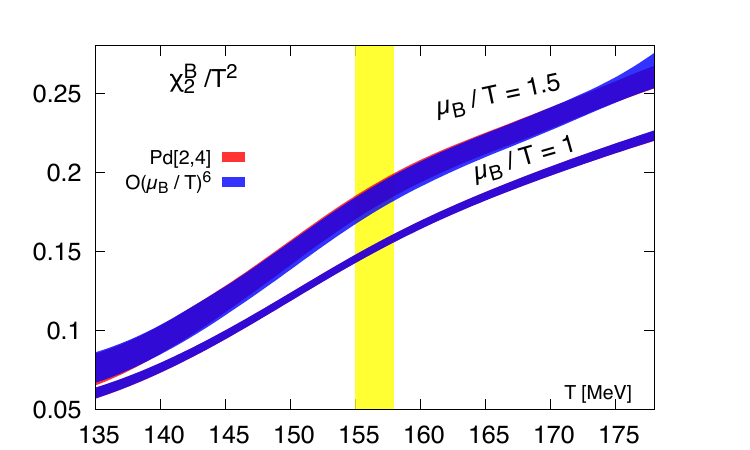}
\includegraphics[scale=0.59]{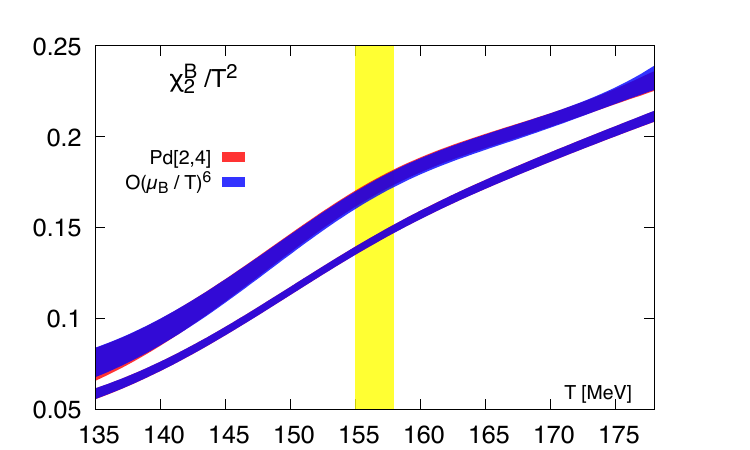}
\end{center}
\caption{Comparison of Taylor expansions and [n,4] {\pade} approximants for the pressure ($n=4$) (top), net baryon-number density ($n=3$) (middle) and the second order cumulant ($n=2$) of net baryon-number fluctuations (bottom) versus temperature
for several values of the baryon chemical potential. 
Shown are results for the
cases $\mu_Q=\mu_S=0$ (left) and 
$\ \mu_Q=0,n_S=0$,(right), respectively.
The Taylor expansions are based on the continuum and spline interpolated data shown in Fig.~\ref{fig:taylor8}.
}
\label{fig:EoS2}
\end{figure*}

In the entire temperature interval 
$135~{\rm MeV}\le T\le  165~{\rm MeV}$ the expansion 
coefficient $\cb_0^{B,8}$ is negative for $\mu_Q=\mu_S=0$ as well as for $\mu_Q=0$, $n_S=0$. 
It will dominate 
the expansion at large $\hmu_B$ and thus forces 
the Taylor expansion of $P/T^4$ to have a maximum at
some value of $\hmu_B^{\rm max}$. As the net baryon-number density is the derivative of $P/T^4$ with respect
to $\hmu_B$ it has a maximum below $\hmu_B^{\rm max}$ and
vanishes at $\hmu_B^{\rm max}$. Similarly, the 
second order cumulant reaches a maximum at an even smaller value of $\hmu_B$.
As can be seen in Fig.~\ref{fig:padetaylor8} the [n,4] Pad\'e approximants (blue bands), the direct $\hmu_B$-derivative of the [4,4] Pad\'e (green bands) and the Taylor expansions (red bands) agree quite well up to values of the chemical potentials
close to the respective value of $\hmu_B^{\rm max}$ and this maximum arises at larger $\hmu_B$ as the
temperature increases. This is in accordance with the
increase of the estimator $\mu_{c,4}^{MR} (T)$ for 
the magnitude of the Pad\'e poles given in Fig.~\ref{fig:dist}.

Starting with a Taylor series limited to $8^{\rm th}$ order obviously the expansions possible for higher order
derivatives becomes shorter. Correspondingly, the
order of a [n,4] Pad\'e used by us becomes smaller. If,
however, the Pad\'e approximant used for the original 
pressure series, {\it i.e.} in our case the [4,4]
Pad\'e approximant, provides a good approximation for
the pressure in the complex $\hmu_B$-plane, taking
directly subsequent derivatives with respect to $\hmu_B$
will give good resummed approximants for {\it e.g.}
the net baryon-number density and higher order cumulants. In Fig.~\ref{fig:padetaylor8} we thus also
show approximations for $n_B/T^3$ and $\chi_2^B$
obtained by taking the first and second derivative 
of the [4,4] Pad\'e approximant of $P/T^4$ (green bands). By construction the poles of these approximants
are identical to those of the [4,4] Pad\'e approximant of $P/T^4$. As can be seen these derivatives agree with 
the corresponding [n,4] approximants up to values of
$\hmu_B$ similar to those where the latter start to 
differ from the corresponding Taylor series. 

Although the radius of convergence for the Taylor series 
of all higher order cumulants is determined by that of the
pressure series, the currently available $8^{\rm th}$ order
Taylor series for the pressure clearly does provide a reliable
approximation for higher order cumulants only in a smaller 
interval of $\hmu_B$ values. We consider the range of $\mu_B$ values
indicated by the range of error bands given in Fig.~\ref{fig:padetaylor8} as the region where current results
on the pressure, net baryon-number density and the second order
cumulant of net baryon-number fluctuations are reliable. As can
be seen in that figure this range of baryon chemical potentials 
is somewhat larger at higher temperatures than at 
lower temperatures.

In Fig.~\ref{fig:EoS2} we compare results obtained for these observables using Taylor expansions as well as Pad\'e approximants
for several values of $\hmu_B$. We show results in the entire 
temperature range $135~{\rm MeV}\le T \le 175~{\rm MeV}$ using values of the chemical potential up to the 
largest value indicated by the bands given in
Fig.~\ref{fig:padetaylor8}. 
As can be seen for the pressure we find
excellent agreement up to values of the chemical potential $\hmu_B\simeq 2.5$. The corresponding largest
values of $\hmu_B$ for $n_B/T^3$ and $\chi_2^B$ are
$\hmu_B=2$ and $1.5$, respectively. This choice of 
$\hmu_B$ values is enforced by demanding good agreement
between Taylor series results and Pad\'e approximants 
at the lowest temperature. At higher temperatures Figs.~\ref{fig:dist}
and Fig.~\ref{fig:padetaylor8} suggest that 
in the vicinity of $T_{pc}$ the range of $\hmu_B$ values, in which $8^{\rm th}$ order Taylor series can provide reliable results is larger, e.g. $\hmu_B \lsim 3$ for $P/T^4$. 

\section{Conclusions}

We have presented results for eighth order  Taylor series expansions of the pressure in (2+1)-flavor QCD for isospin symmetric matter corresponding to vanishing electric charge chemical potentials. From this Taylor series 
we derived the first two cumulants of net baryon-number fluctuations, corresponding to the mean and  variance 
of the net baryon-number distribution.
We used Pad\'e approximants to resum these Taylor series.

We have shown that the [4,4] Pad\'e
approximant, which reproduces the 
eighth order Taylor series of the 
pressure series has only complex 
poles in the entire temperature 
interval $135~{\rm MeV}\le T\le 165~{\rm MeV}$, which gives further support to
the observation that a possible critical point in the QCD phase diagram may only be found at temperatures below $135~{\rm MeV}$. From the location of the poles in the complex plane we estimate the radius of the 
convergence for these Taylor series expansions to be slightly temperature dependent, increasing from $\hmu_{B,c}\simeq 2.2$ at $T=135~{\rm MeV}$ to $\hmu_{B,c}\simeq 3.2$ at $T=165~{\rm MeV}$. 
In the vicinity of the pseudo-critical temperature, $T_{pc}\simeq 156.5$~MeV, we find $\mu_B/T\ \gsim\ 2.9$ at vanishing strangeness chemical potential and somewhat larger values for strangeness neutral matter. 
If poles of [n,n] Pad\'e
approximants continue to lie in the complex 
$\hmu_B$-plane an efficient resummation of the 
Taylor series of the QCD pressure to even larger values of
$\hmu_B$ will be possible in this temperature range. 

A comparison of Taylor series and Pad\'e
approximants for the Taylor series of the pressure and the first two cumulants of 
net baryon-number fluctuations allows us 
to estimate the range of $\hmu_B$ values in which current series expansions give reliable results. For the pressure and the first two cumulants in (2+1)-flavor QCD we deduce that 
the current eighth order series for the pressure and its derivatives agree well with
the resummed [4,4] Pad\'e approximants and its
derivatives for $\hmu_B\le 2.5$ (pressure),
$2.0$ (net number-density) and $1.5$ (second
order cumulant). All data presented in the figures of this paper can be
found in~\cite{pubdata}.

\vspace{0.5cm}
\emph{Acknowledgments.---} 
This work was supported by: (i) The U.S. Department of Energy, Office of
Science, Office of Nuclear Physics through the Contract No. DE-SC0012704;
(ii) The U.S. Department of Energy, Office of Science, Office of Nuclear
Physics and Office of Advanced Scientific Computing Research within the
framework of Scientific Discovery through Advance Computing (SciDAC) award
{\it Computing the Properties of Matter with Leadership Computing Resources};
(iii) The Deutsche Forschungsgemeinschaft (DFG, German Research Foundation) - Project number 315477589-TRR 211;
(iv) The grant 05P2018 (ErUM-FSP T01) of the German Bundesministerium f\"ur Bildung und Forschung;
(v) The grant 283286 of the European Union.
D.B.was supported by the Intel Corporation.

This research used awards of computer time provided by:
(i) The INCITE program at Oak Ridge Leadership Computing Facility, a DOE
Office of Science User Facility operated under Contract No. DE-AC05-00OR22725;
(ii) The ALCC program at National Energy Research Scientific Computing Center,
a U.S. Department of Energy Office of Science User Facility operated under
Contract No. DE-AC02-05CH11231;
(iii) The INCITE program at Argonne Leadership Computing Facility, a U.S.
Department of Energy Office of Science User Facility operated under Contract
No. DE-AC02-06CH11357;
(iv) The USQCD resources at the Thomas Jefferson National Accelerator Facility.

This research also used computing resources made available through:
(i) a  PRACE grant at CINECA, Italy;
(ii) the Gauss Center at NIC-J\"ulich, Germany;
(iii) the GPU-cluster at Bielefeld University, Germany.
\vspace{0.2cm}

\appendix
\allowdisplaybreaks
\section{Taylor expansion coefficients for strangeness neutral, isospin symmetric QCD matter}
\label{app:Pn}
We give here the general form of the eighth order
expansion coefficients, $\cb_n^{B,k}$ for $n^{\rm th}$
order cumulants of net baryon-number fluctuations.
In the context of this work it only is needed for
the pressure series ($n=0$) in the case $\hmu_Q = 0$, which corresponds to setting $q_n=0$ in the following expression.
Expansion coefficients of all other cumulants, $\cb_n^{B,k}$ that involve only cumulants $\chi_{ijk}^{BQS}$ with $i+j+k\le 8$ are given in Appendix A of \cite{Bazavov:2020bjn}.
\begin{widetext}
\begin{eqnarray}
\bar{\chi}^{B,8}_n&=& 40320 \chi^{BQS}_{n02} s_{1} s_{7} + 40320 \chi^{BQS}_{n02} s_{3} s_{5} + 6720  \chi^{BQS}_{n04} s_{1}^{3} s_{5} +  10080 \chi^{BQS}_{n04} s_{1}^{2} s_{3}^{2}  + 336 \chi^{BQS}_{n06} s_{1}^{5} s_{3} \nonumber \\
&+&  \chi^{BQS}_{n08} s_{1}^{8}
 + 40320 \chi^{BQS}_{n11} q_{1} s_{7} + 40320 \chi^{BQS}_{n11} q_{3} s_{5} + 40320 \chi^{BQS}_{n11} q_{5} s_{3}
 + 40320 \chi^{BQS}_{n11} q_{7} s_{1} + 20160 \chi^{BQS}_{n13} q_{1} s_{1}^{2} s_{5} \nonumber \\
 &+& 20160 \chi^{BQS}_{n13} q_{1} s_{1} s_{3}^{2} 
 + 20160 \chi^{BQS}_{n13} q_{3} s_{1}^{2} s_{3} + 6720 \chi^{BQS}_{n13} q_{5} s_{1}^{3} + 1680 \chi^{BQS}_{n15} q_{1} s_{1}^{4} s_{3} + 
 + 336 \chi^{BQS}_{n15} q_{3} s_{1}^{5} + 8 \chi^{BQS}_{n17} q_{1} s_{1}^{7} \nonumber \\
 &+& 40320 \chi^{BQS}_{n20} q_{1} q_{7}
 + 40320 \chi^{BQS}_{n20} q_{3} q_{5} + 20160 \chi^{BQS}_{n22} q_{1}^{2} s_{1} s_{5} + 10080 \chi^{BQS}_{n22} q_{1}^{2} s_{3}^{2} 
 + 40320 \chi^{BQS}_{n22} q_{1} q_{3} s_{1} s_{3} \nonumber \\
 &+& 20160 \chi^{BQS}_{n22} q_{1} q_{5} s_{1}^{2} + 10080 \chi^{BQS}_{n22} q_{3}^{2} s_{1}^{2} 
 + 3360 \chi^{BQS}_{n24} q_{1}^{2} s_{1}^{3} s_{3} + 1680 \chi^{BQS}_{n24} q_{1} q_{3} s_{1}^{4} + 28 \chi^{BQS}_{n26} q_{1}^{2} s_{1}^{6} 
 + 6720 \chi^{BQS}_{n31} q_{1}^{3} s_{5} \nonumber \\
 &+& 20160 \chi^{BQS}_{n31} q_{1}^{2} q_{3} s_{3} + 20160 \chi^{BQS}_{n31} q_{1}^{2} q_{5} s_{1}
 + 20160 \chi^{BQS}_{n31} q_{1} q_{3}^{2} s_{1} + 3360 \chi^{BQS}_{n33} q_{1}^{3} s_{1}^{2} s_{3} + 3360 \chi^{BQS}_{n33} q_{1}^{2} q_{3} s_{1}^{3} \nonumber \\
 &+& 56 \chi^{BQS}_{n35} q_{1}^{3} s_{1}^{5} 
 + 6720 \chi^{BQS}_{n40} q_{1}^{3} q_{5} + 10080 \chi^{BQS}_{n40} q_{1}^{2} q_{3}^{2}
 + 1680 \chi^{BQS}_{n42} q_{1}^{4} s_{1} s_{3} + 3360 \chi^{BQS}_{n42} q_{1}^{3} q_{3} s_{1}^{2} + 70 \chi^{BQS}_{n44} q_{1}^{4} s_{1}^{4} \nonumber \\
 &+& 336 \chi^{BQS}_{n51} q_{1}^{5} s_{3} + 1680 \chi^{BQS}_{n51} q_{1}^{4} q_{3} s_{1} + 56 \chi^{BQS}_{n53} q_{1}^{5} s_{1}^{3} 
 + 336 \chi^{BQS}_{n60} q_{1}^{5} q_{3} + 28 \chi^{BQS}_{n62} q_{1}^{6} s_{1}^{2}+ 8 \chi^{BQS}_{n71} q_{1}^{7} s_{1} \nonumber \\
 &+& \chi^{BQS}_{n80} q_{1}^{8} 
 + 40320 \chi^{BQS}_{n+101} s_{7} + 20160 \chi^{BQS}_{n+103} s_{1}^{2} s_{5} + 20160 \chi^{BQS}_{n+103} s_{1} s_{3}^{2} + 1680 \chi^{BQS}_{n+105} s_{1}^{4} s_{3} + 8 \chi^{BQS}_{n+107} s_{1}^{7} \nonumber \\
 &+& 40320 \chi^{BQS}_{n+110} q_{7}  
 + 40320 \chi^{BQS}_{n+112} q_{1} s_{1} s_{5} + 20160 \chi^{BQS}_{n+112} q_{1} s_{3}^{2} + 40320 \chi^{BQS}_{n+112} q_{3} s_{1} s_{3} 
 + 20160 \chi^{BQS}_{n+112} q_{5} s_{1}^{2} \nonumber \\
 &+& 6720 \chi^{BQS}_{n+114} q_{1} s_{1}^{3} s_{3} + 1680 \chi^{BQS}_{n+114} q_{3} s_{1}^{4} 
 + 56 \chi^{BQS}_{n+116} q_{1} s_{1}^{6} + 20160 \chi^{BQS}_{n+121} q_{1}^{2} s_{5} + 40320 \chi^{BQS}_{n+121} q_{1} q_{3} s_{3} \nonumber \\
 &+& 40320 \chi^{BQS}_{n+121} q_{1} q_{5} s_{1} + 20160 \chi^{BQS}_{n+121} q_{3}^{2} s_{1} + 10080 \chi^{BQS}_{n+123} q_{1}^{2} s_{1}^{2} s_{3} 
 + 6720 \chi^{BQS}_{n+123} q_{1} q_{3} s_{1}^{3} + 168 \chi^{BQS}_{n+125} q_{1}^{2} s_{1}^{5} \nonumber \\
 &+& 20160 \chi^{BQS}_{n+130} q_{1}^{2} q_{5} 
 + 20160 \chi^{BQS}_{n+130} q_{1} q_{3}^{2} + 6720 \chi^{BQS}_{n+132} q_{1}^{3} s_{1} s_{3} + 10080 \chi^{BQS}_{n+132} q_{1}^{2} q_{3} s_{1}^{2} 
 + 280 \chi^{BQS}_{n+134} q_{1}^{3} s_{1}^{4} \nonumber \\
 &+& 1680 \chi^{BQS}_{141} q_{1}^{4} s_{3} + 6720 \chi^{BQS}_{141} q_{1}^{3} q_{3} s_{1}
 + 280 \chi^{BQS}_{n+143} q_{1}^{4} s_{1}^{3} + 1680 \chi^{BQS}_{150} q_{1}^{4} q_{3} + 168 \chi^{BQS}_{152} q_{1}^{5} s_{1}^{2} 
 + 56 \chi^{BQS}_{n+161} q_{1}^{6} s_{1} \nonumber \\
 &+& 8 \chi^{BQS}_{n+170} q_{1}^{7} + 20160 \chi^{BQS}_{n+202} s_{1} s_{5}
 + 10080 \chi^{BQS}_{n+202} s_{3}^{2} + 3360 \chi^{BQS}_{n+204} s_{1}^{3} s_{3} + 28 \chi^{BQS}_{n+206} s_{1}^{6} 
+ 20160 \chi^{BQS}_{n+211} q_{1} s_{5} \nonumber \\
&+& 20160 \chi^{BQS}_{n+211} q_{3} s_{3} + 20160 \chi^{BQS}_{n+211} q_{5} s_{1} 
+ 10080 \chi^{BQS}_{n+213} q_{1} s_{1}^{2} s_{3} + 3360 \chi^{BQS}_{n+213} q_{3} s_{1}^{3} + 168 \chi^{BQS}_{n+215} q_{1} s_{1}^{5} \nonumber \\
&+& 20160 \chi^{BQS}_{n+220} q_{1} q_{5} + 10080 \chi^{BQS}_{n+220} q_{3}^{2} + 10080 \chi^{BQS}_{n+222} q_{1}^{2} s_{1} s_{3} 
+ 10080 \chi^{BQS}_{n+222} q_{1} q_{3} s_{1}^{2} + 420 \chi^{BQS}_{n+224} q_{1}^{2} s_{1}^{4} \nonumber \\
&+& 3360 \chi^{BQS}_{n+231} q_{1}^{3} s_{3}
 + 10080 \chi^{BQS}_{n+231} q_{1}^{2} q_{3} s_{1} + 560 \chi^{BQS}_{n+233} q_{1}^{3} s_{1}^{3} + 3360 \chi^{BQS}_{n+240} q_{1}^{3} q_{3}
 + 420 \chi^{BQS}_{n+242} q_{1}^{4} s_{1}^{2} + 168 \chi^{BQS}_{n+251} q_{1}^{5} s_{1} \nonumber \\
 &+& 28 \chi^{BQS}_{n+260} q_{1}^{6}
 + 6720 \chi^{BQS}_{n+301} s_{5} + 3360 \chi^{BQS}_{n+303} s_{1}^{2} s_{3} + 56 \chi^{BQS}_{n+305} s_{1}^{5}
 + 6720 \chi^{BQS}_{n+310} q_{5} + 6720 \chi^{BQS}_{n+312} q_{1} s_{1} s_{3} \nonumber \\
 &+& 3360 \chi^{BQS}_{n+312} q_{3} s_{1}^{2} 
 + 280 \chi^{BQS}_{n+314} q_{1} s_{1}^{4} + 3360 \chi^{BQS}_{n+321} q_{1}^{2} s_{3} + 6720 \chi^{BQS}_{n+321} q_{1} q_{3} s_{1} 
 + 560 \chi^{BQS}_{n+323} q_{1}^{2} s_{1}^{3} + 3360 \chi^{BQS}_{n+330} q_{1}^{2} q_{3} \nonumber \\
 &+& 560 \chi^{BQS}_{n+332} q_{1}^{3} s_{1}^{2} 
 + 280 \chi^{BQS}_{n+341} q_{1}^{4} s_{1} + 56 \chi^{BQS}_{n+350} q_{1}^{5} + 1680 \chi^{BQS}_{n+402} s_{1} s_{3} 
 + 70 \chi^{BQS}_{n+404} s_{1}^{4} + 1680 \chi^{BQS}_{n+411} q_{1} s_{3} \nonumber \\
 &+& 1680 \chi^{BQS}_{n+411} q_{3} s_{1} 
 + 280 \chi^{BQS}_{n+413} q_{1} s_{1}^{3} + 1680 \chi^{BQS}_{n+420} q_{1} q_{3} + 420 \chi^{BQS}_{n+422} q_{1}^{2} s_{1}^{2}
 + 280 \chi^{BQS}_{n+431} q_{1}^{3} s_{1} + 70 \chi^{BQS}_{n+440} q_{1}^{4} \nonumber \\
 &+& 336 \chi^{BQS}_{n+501} s_{3} + 56 \chi^{BQS}_{n+503} s_{1}^{3} 
 + 336 \chi^{BQS}_{n+510} q_{3} + 168 \chi^{BQS}_{n+512} q_{1} s_{1}^{2} + 168 \chi^{BQS}_{n+521} q_{1}^{2} s_{1} 
 + 56 \chi^{BQS}_{n+530} q_{1}^{3} + 28 \chi^{BQS}_{n+602} s_{1}^{2}  \nonumber \\
 &+& 56 \chi^{BQS}_{n+611} q_{1} s_{1} 
 + 28 \chi^{BQS}_{n+620} q_{1}^{2} + 8 \chi^{BQS}_{n+701} s_{1} + 8 \chi^{BQS}_{n+710} q_{1} + \chi^{BQS}_{n+800} \nonumber \\
\end{eqnarray}

\end{widetext}

\begin{figure*}[t]
\includegraphics[scale=0.48]{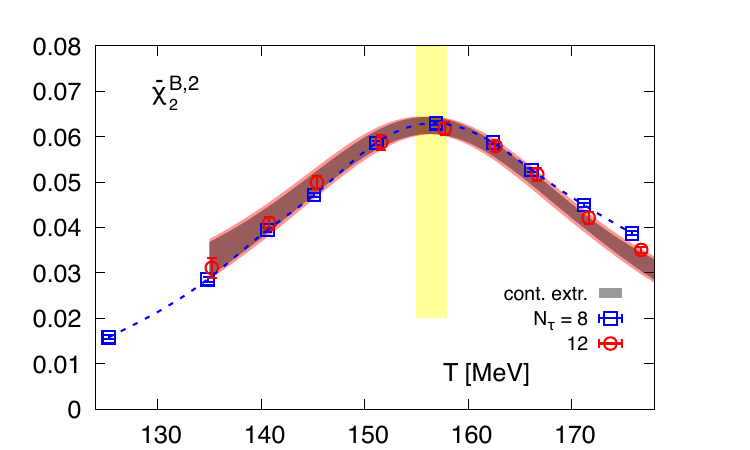}
\hspace*{-0.4cm}
\includegraphics[scale=0.48]{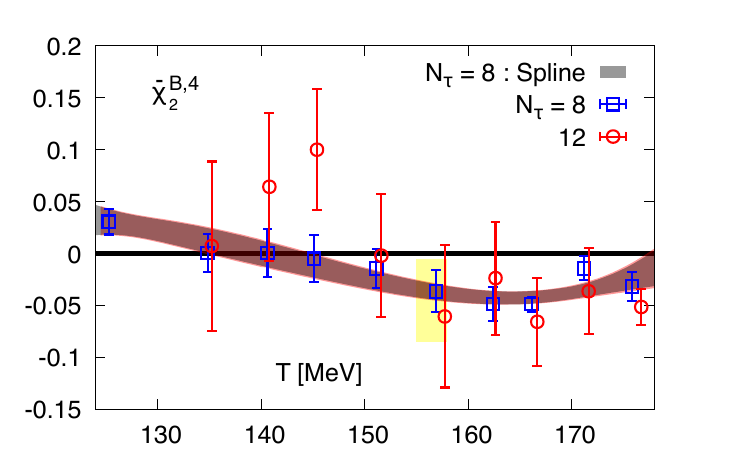}
\hspace*{-0.4cm}
\includegraphics[scale=0.48]{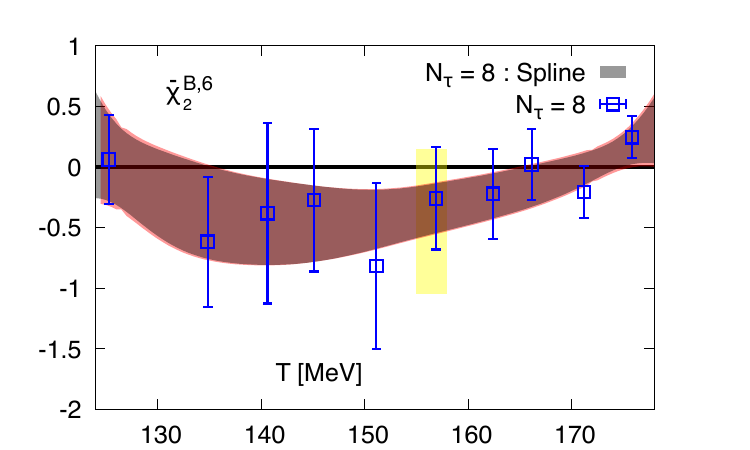}
\caption{Taylor expansion coefficients, $\cb_2^{B,k}$, of the second order 
cumulant of net baryon-number fluctuations in
(2+1)-flavor QCD. 
}
\label{fig:taylor8b}
\end{figure*}

\section{Taylor expansion coefficients
for the \boldmath$\cb_2^B$ in the case $\mu_Q=0$, $n_S=0$}
\label{app:chi2B}

In Fig.~\ref{fig:taylor8b} we show the expansion coefficients, $\cb_2^{B,k}$, for the Taylor series of
the second order cumulant of net baryon-number fluctuations in
(2+1)-flavor QCD defined in Eq.~\ref{chineutral}.

\section{Location of real and imaginary poles
in the parameter space \boldmath($c_{6,2},c_{8,2}$)}
\label{app:zpzm}
We discuss here the occurrence of complex and real
poles in the plane of the two real expansion parameters appearing in the Taylor series for the pressure, 
($c_{6,2},c_{8,2}$), and characterize the location 
of [4,4] Pad\'e approximants constructed from the $8^{th}$ order Taylor series of the pressure in (2+1)-flavor QCD.

\begin{figure}[t]
	\begin{center}
\includegraphics[scale=0.65]{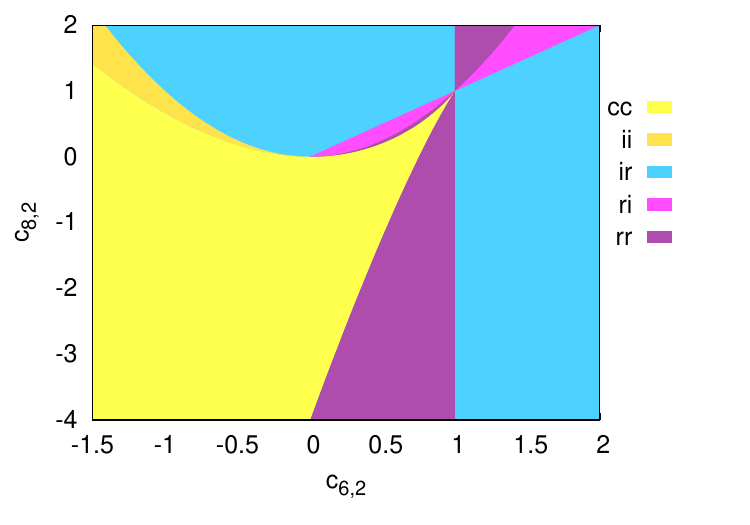}
\end{center}
\vspace{-0.5cm}
\caption{Location of poles in
the complex $\hmu_B$-plane as function of the couplings $c_{6,2}$
and $c_{8,2}$. In the yellow area the four poles are
complex (cc) with ${\rm Re}\mu_B \ne 0$ and ${\rm Im}\mu_B \ne 0$. In the other regions they come as pairs of two purely real (r) or purely imaginary (i) poles.
The notation $xy$ in the legend of the figure indicates that there
is a pair of poles of type $x$ and another pair of type $y$ where the poles of type $x$ are closest to the origin.}
\label{fig:poles}
\end{figure}

As discussed in Section~\ref{radius}
there is a triangular shaped region in this parameter space, bounded by lines $c_{8,2}^\pm$
given in Eq.~\ref{c8pm}, in which all four poles of the
[4,4] Pad\'e are complex with non-vanishing real and imaginary parts. 
Outside this region  poles of the [4,4] Pad\'e are either real or purely imaginary.
For $z^\pm>0$ there exists a pair of real poles in terms of $\hmu_B$, for $z^\pm<0$ one has a pair of 
purely imaginary poles. In the parameter space 
($c_{6,2},c_{8,2}$) one can have two pairs of 
purely imaginary poles ($ii$), two pairs of 
real poles ($rr$), or a pair of each of these 
types, ($ir$) or ($ri$). In the latter case we 
use the convention that the first letter corresponds to that pair of poles that is closest 
to the origin. The parameters ($c_{6,2},c_{8,2}$)  for which these different types of poles appear are shown in Fig.~\ref{fig:poles}.

In the following we give some further details on the boundaries for the different regions in parameter space:
We re-write Eq.~\ref{zeroes} as
\begin{eqnarray}
z^\pm &=&\frac{c_{8,2} - c_{6,2}}{2 (c_{8,2} -c_{6,2}^2)} \pm \nonumber \\ 
&&\frac{\sqrt{(c_{8,2} - c_{6,2})^2 +4
(c_{8,2} -c_{6,2}^2) (1-c_{6,2})}}{2 (c_{8,2} -c_{6,2}^2)}\; .
\label{apzeroes}
\end{eqnarray}

The zeroes $z^+$ and $z^-$ are related to each other through
\begin{equation}
    z^+ z^- = \frac{1-c_{6,2}}{c_{6,2}^2 - c_{8,2}}
    \;\; .
    \label{zsign}
\end{equation}
Outside the region bounded by $c_{8,2}^\pm$ both zeroes have the same sign,
if $z^+ z^- >0$, {\it i.e.} if
numerator and denominator in
Eq.~\ref{zsign} have the same sign, which is the case for
\begin{eqnarray}
   {\rm either}\; c_{6,2} > 1  &\;\;{\rm and}\;\;& c_{8,2} > c_{6,2}^2 \; , \label{plus1} \\[1mm]
   {\rm or}\;\; 
   c_{6,2} < 1 &\;\;{\rm and}\;\;& c_{8,2} <
   c_{6,2}^2\; . \label{plus2}
\end{eqnarray}
In the first case it is obvious that
$c_{8,2} > c_{6,2}^2 >c_{6,2}$ holds.
It thus is evident from Eq.~\ref{apzeroes} that $z^+ >0$ and the region defined in Eq.~\ref{plus1} corresponds to a region with two real poles in
the complex $\hmu_B$-plane. For all other regions 
with $c_{6,2} >1$ a pair of real and a pair of purely 
imaginary poles exists. However, only for $c_{8,2} < c_{6,2}$ it is the imaginary pair of poles that is closest to the origin.

In the second case, Eq.~\ref{plus2}, we obtain from Eq.~\ref{c8pm},  
\begin{equation}
 c_{6,2}^2 - c_{8,2}^+ =  (1-c_{6,2}) (2- c_{6,2} -2 \sqrt{1-c_{6,2}}) > 0\; .
\end{equation}
It thus is evident from Eq.~\ref{apzeroes}
that $z^+<0$ for $c_{6,2}<0$. In the range
$c_{8,2}^+<c_{8,2}<c_{6,2}^2$ one thus finds
two pairs of purely imaginary poles in the
complex $\hmu_B$-plane. On the other hand,
for $0<c_{6,2}<1$ one finds in the same $c_{8,2}$ interval that $z^->0$. In this case
one thus has two pairs of real poles in the
complex $\hmu_B$-plane.
Finally there is a second region for purely 
real poles, which is allowed by Eq.~\ref{plus2}. This is the case of $c_{8,2}<c_{8,2}^-$, as  also in this case one finds $z^->0$. 
In all other case one finds a pair of real and
a pair of complex poles. These different parameter regions in the $(c_{6,2},c_{8,2})$
plane are shown in Fig.~\ref{fig:poles}.

\bibliography{bibliography}
\end{document}